\RequirePackage{rotating}

\documentclass[fleqn,usenatbib]{mnras}

\usepackage{newtxtext,newtxmath}
\usepackage{supertabular}

\usepackage[T1]{fontenc}

\usepackage{subcaption}
\usepackage{academicons}

\DeclareRobustCommand{\VAN}[3]{#2}
\let\VANthebibliography\thebibliography
\def\thebibliography{\DeclareRobustCommand{\VAN}[3]{##3}\VANthebibliography}

\usepackage{graphicx}	
\usepackage{amsmath}	

\usepackage{rotating}
\usepackage{bm}
\usepackage{longtable,tabu}
\usepackage{xcolor}
\usepackage{pifont}
\usepackage{soul}
\usepackage{ulem}
\usepackage{caption}

\usepackage{graphicx,pifont}
\let\oldding\ding
\renewcommand{\ding}[2][1]{\scalebox{#1}{\oldding{#2}}}

\usepackage{array}
\newcolumntype{H}{>{\setbox0=\hbox\bgroup}c<{\egroup}@{}}

\newcommand{\gm}{\textbf}

\definecolor{orcidlogocol}{HTML}{A6CE39}

\title[RR Lyrae in the HOWVAST survey]{Taking the pulse of the outer Milky Way with HOWVAST: an RR Lyrae density profile out to >200 kpc}

\author[Medina et al.]{
Gustavo E. Medina{$^{\href{https://orcid.org/0000-0003-0105-9576}{\includegraphics[scale=.035]{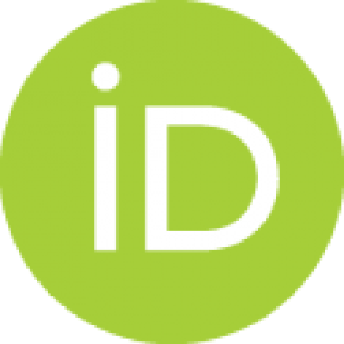}}}$},$^{1,2}$\thanks{E-mail: gustavo.medina@astro.utoronto.ca} 
Ricardo R. Mu\~{n}oz{$^{\href{https://orcid.org/0000-0002-0810-5558}{\includegraphics[scale=.035]{orcid-ID.eps}}}$},$^{3}$ 
Jeffrey L. Carlin{$^{\href{https://orcid.org/0000-0002-3936-9628}{\includegraphics[scale=.035]{orcid-ID.eps}}}$},$^{4}$ 
A. Katherina Vivas{$^{\href{https://orcid.org/0000-0003-4341-6172}{\includegraphics[scale=.035]{orcid-ID.eps}}}$},$^{5}$ 
Eva K. Grebel,$^{2}$ 
\newauthor
Clara E. Mart\'inez-V\'azquez{$^{\href{https://orcid.org/0000-0002-9144-7726}{\includegraphics[scale=.035]{orcid-ID.eps}}}$},$^{6}$ 
and 
Camilla J. Hansen{$^{\href{https://orcid.org/0000-0002-7277-7922}{\includegraphics[scale=.035]{orcid-ID.eps}}}$}$^{7}$ 
\\
$^{1}$David A. Dunlap Department of Astronomy \& Astrophysics, University of Toronto, 50 St George Street, Toronto ON M5S 3H4, Canada\\
$^{2}$Astronomisches Rechen-Institut, Zentrum f{\"u}r Astronomie der Universit{\"a}t Heidelberg, M{\"o}nchhofstr. 12-14, 69120 Heidelberg, Germany\\
$^{3}$Departamento de Astronom\'ia, Universidad de Chile, Camino El Observatorio 1515, Las Condes, Santiago, Chile\\
$^{4}$AURA/Rubin Observatory Project Office, 950 North Cherry Avenue, Tucson, AZ 85719, USA\\
$^{5}$Cerro Tololo Inter-American Observatory/NSF's NOIRLab, Casilla 603, La Serena, Chile\\
$^{6}$Gemini Observatory/NSF's NOIRLab, 670 N. A'ohoku Place, Hilo, HI 96720, USA\\
$^{7}$Goethe University Frankfurt, Institute for Applied Physics, Max-von-Laue Str. 11, 60438 Frankfurt am Main, Germany\\
}

\date{Accepted XXX. Received YYY; in original form ZZZ}

\pubyear{2024}

\begin{document}
\label{firstpage}
\pagerange{\pageref{firstpage}--\pageref{lastpage}}
\maketitle

\begin{abstract}

In order to constrain the evolutionary history of the Milky Way, we hunt for faint RR Lyrae stars (RRLs) using Dark Energy Camera data from the High cadence Transient Survey (HiTS) and the Halo Outskirts With Variable Stars (HOWVAST) survey. 
We report the detection of $\sim$500 RRLs, including previously identified stars and $\sim$90 RRLs not yet reported. 
We identify 9 new RRLs beyond $100$\,kpc from the Sun, most of which are classified as fundamental-mode pulsators. 
The periods and amplitudes of the distant RRLs do not place them in either one of the two classical Oosterhoff groups, but in the Oosterhoff intermediate region. 
We detect two groups of clumped distant RRLs with similar distances and equatorial coordinates, which we interpret as an indication of their association with undiscovered bound or unbound satellites.
We study the halo density profile using spheroidal and ellipsoidal ($q=0.7$) models, following a Markov chain Monte Carlo methodology. 
For a spheroidal halo, our derived radial profile is consistent with a broken power-law with a break at $18.1^{+2.1}_{-1.1}$\,kpc separating the inner and the outer halo, and an outer slope of  $-4.47^{+0.11}_{-0.18}$. 
For an ellipsoidal halo, the break is located at $24.3^{+2.6}_{-3.2}$\,kpc and the outer slope is $-4.57^{+0.17}_{-0.25}$. 
The break in the density profile is a feature visible in different directions of the halo.
The similarity of these radial distributions with previous values reported in the literature seems to depend on the regions of the sky surveyed (direction and total area) and halo tracer used. 
Our findings are compatible with simulations and observations that predict that the outer regions of Milky Way-like galaxies are mainly composed of accreted material.

\end{abstract}





\section{Introduction}
\label{sec:intro}

In the currently favoured cosmological framework, the $\Lambda$ cold dark matter ($\Lambda$-CDM) model, galaxies assemble hierarchically through the accretion of smaller systems.
The Milky Way (MW) and similar massive disc  galaxies likely experienced numerous mergers in their early history as part of their hierarchical formation \citep[see, e.g.,][]{Press1974,Blumenthal1984,BJ05,Montalban2021}.
The stellar halos of these galaxies provide key information to help reconstruct their formation conditions. 
For the MW, in particular, compelling evidence for 
past and ongoing accretion events have been identified in present-day inner and outer halo stellar populations, unveiling details of gravitational interactions with massive satellites such as the Sagittarius stream \citep[e.g.,][]{Ibata1994,Majewski2003,Vivas2006}, 
Gaia-Sausage-Enceladus \citep[GSE; e.g.,][]{Belokurov2018b,Helmi2018,Haywood2018}, and the infall of the Magellanic Clouds \citep[e.g.,][]{Mathewson1974,Besla2007,Zaritsky2020,Erkal2021}.

The accretion history of a particular halo is also imprinted in the shape of its stellar density profile \citep[e.g.,][]{BJ05,Cooper2013}, as the mass distribution is sensitive to properties such as the halo formation time, the amount of stellar mass accreted, and how long ago the last mergers took place \citep[see e.g.][]{Pillepich2014}. 
The slope of the number density profile of outer halo stars, in particular, has been shown to be a parameter of cosmological significance, closely related to the accretion history of MW-like galaxies \citep{Juric2008,Pillepich2014,Merritt2016,Slater2016}.

However, well-characterized MW stars with precisely determined distances and reliable classifications at large distances are rare, especially close to the `edge' of the MW \citep[$292 \pm 61$\,kpc, when defined as the point at which virialized material has completed at least two pericentric passages;][]{Deason2020}. 
Among the commonly used stellar tracers are the RR Lyrae stars \citep[RRLs; see e.g.][]{Dra13_100kpc, Medina2018,Stringer21,Huang2022}. 
RR Lyrae variables are old (ages $>$ 10\,Gyr) and metal-poor ([Fe/H] typically $< -0.5$) stars of paramount importance for Galactic studies given their status of precise distance indicators \citep[$\sim$5\,per\,cent from period-luminosity relations;][and references therein]{Catelan2015}, and their easy identification in time-domain surveys (given their intrinsically high luminosity and light curve shapes). 
The pulsation periods of RRLs typically range from 0.2 to 0.9\,d \citep[e.g.,][]{Smith1995} and they are mainly classified into three subtypes, according to the nature of the pulsations, their periods, amplitudes, and light curve shapes.
RRab stars are fundamental-mode pulsators with saw-tooth light curves and typically longer periods, whereas RRc stars pulsate in the first overtone and have sinusoidal light curves with shorter periods.  
RRd stars are a less common subtype and pulsate in the fundamental mode and the first overtone simultaneously, with the first overtone being the principal mode of pulsation. 

Given that RRLs are ubiquitous in the halo and dwarf galaxies \citep[see e.g.][and references therein]{Martinez2019}, they are typically used for numerous astrophysical applications.
For instance, RRLs are useful to shed light onto the genesis of the stellar halo \citep[see e.g.][]{Catelan2004,Viv04,Vivas2006,Hansen2011,Fiorentino2015,Torrealba2015,Hansen2016,Belokurov2018a,Deason2017,Dekany2018,Hernit2018,Mateu2018,Prudil2019,Prudil2021,Monelli2022,Medina2023}. 
Furthermore, \citet{Ses14} proposed that RRLs can be used as tracers of yet undiscovered low luminosity satellites, and \citet{BakerWillman2015} suggested that even small groups of halo RRLs can serve for this purpose \citep[see e.g.][]{Torrealba2019}.
They can also be used to characterize and find new stellar streams resulting from past accretions events \citep{Vivas01,Duffau2006,Sesar2010,Hendel2018,Iorio2019,PriceWhelan2019,Abbas2021,Prudil2021} and as evidence of the extragalactic origin of overdensities in the disc \citep[e.g.,][]{Mateu2009,PriceWhelan2015}. 
Because RRLs are old, combining their precise distances with proper motions and line-of-sight velocities is key to reconstruct the early assembly of the Galaxy, as notably done for the characterization of the GSE merger event \citep[e.g.,][]{Belokurov2018b, Helmi2018}.

Given that the distances of RRLs are known with great precision, their spatial distribution can be derived and thus they can be used to study the radial density profile of the Galaxy \citep[][]{Wetterer1996,Vivas2006,Cohen2017,Iorio2018}. 
This also makes them excellent tracers of the outermost limits of our Galaxy, as well-characterized stars at such large distances (beyond 100\,kpc) are scarce \citep{Sesar2017,Medina2018,Stringer21}, thus playing a key role in the estimation of the MW mass \citep[see e.g., ][]{Eadie2016,Deason2019,Deason2021,Rodriguez2021,Prudil2022}.

A large number of RRL catalogues have been compiled over the years in existing large sky surveys, which cover a wide range of photometric depths (hence distances) and different regions of the sky. 
These systematic searches include the Quasar Equatorial Survey Team (QUEST) and the La-Silla QUEST surveys \citep{Viv04,Zinn14}, the Northern Sky Variability Survey \citep[NSVS;][]{Kinemuchi2006}, the Sloan Digital Sky Survey (SDSS) Stripe 82 \citep{Sesar2010}, the Catalina surveys \citep{Abbas2014,Dra14,Drake17,Torrealba2015}, the Panoramic Survey Telescope And Rapid Response System survey \citep[Pan-STARRS-1, or PS-1;][]{Chambers2016,Hernitschek2016,Sesar2017}, the Optical Gravitational Lensing Experiment (OGLE) survey \citep[][]{Soszynski2016,Soszynski2019}, the second and third data releases of the {\it Gaia}\ mission (\citealt{Holl2018}, and \citealt{Clementini2019} and \citealt{Clementini2022} using the Specific Objects Study pipeline, SOS), the Dark Energy Survey \citep[DES;][]{DES2016, Stringer21}, and the Zwicky Transient Facility survey \citep[ZTF;][]{Masci2019,Chen2020,Huang2022}.
Only a small subset of these surveys have allowed astronomers to reliably detect RRLs beyond 100\,kpc, mostly due to instrumental limitations, while thousands of RRLs are predicted to be found in the halo between 100 and 300\,kpc \citep{Sanderson2017}. 
In this vein, \citet{Medina2018} identified 16 RRL candidates with Galactocentric distances $>$ 100\,kpc over a $\sim$120\,deg$^2$ area using data from the High cadence Transient Survey \citep[HiTS;][]{Forster16}. 
The current census of distant RRLs is still likely incomplete, and the outer limits of the Galaxy have yet to be comprehensively mapped. 
This serves as motivation for larger surveys focusing on the detection of RRLs at large distances.

In this study we introduce the Halo Outskirts With Variable Stars (HOWVAST) survey, with which we aim to extend the reach of known well-characterized outer halo RRL surveys out to $\sim$270\,kpc. 
In Section~\ref{sec:data}, we describe the survey strategy, the observations carried out for this study, and the methodology followed for data processing. 
In Section~\ref{sec:search}, we provide a detailed description of our RRL selection and classification pipelines, as well as the methods used for the determination of their periods, amplitudes, and distances.  
Additionally, we contrast our detected RRLs with those from the literature and use these comparisons as an indicator of our detection completeness. 
In Section~\ref{sec:100kpc}, we focus our attention on the most distant RRLs in our sample (those with $d_{\rm H} > 100$\,kpc).
Finally, in Section~\ref{sec:density} we study the spatial distribution of our RRLs following a Markov Chain Monte Carlo methodology, and discuss the similarities and differences between our results and studies of other regions of the halo. 
We conclude this manuscript by summarizing our results and outlining the implications of our findings in Section~\ref{sec:discussionSummary}.

\begin{table*}\small
\caption{
Identification numbers and equatorial coordinates of the DECam fields observed by HOWVAST in 2017 and 2018. 
We quote the central J2000 coordinates of our fields and the number of observations per field in the $g$ and $r$ bands (N$_g$ and N$_r$, respectively). Each field covers three square degrees, approximately.  
The fields of the second campaign are labeled according to their positions with respect to the Galactic plane (2018A and 2018B represent the high- and low-Galactic latitude fields).  
}
\label{tab:fields}
\begin{center}

\begin{subtable}{.45\linewidth}
\centering

\begin{tabular}{|c|c|c|c|c|c|}

\hline

Campaign   & Field ID & R.A. & Dec. & N$_g$ & N$_r$ \\

&  & (deg) & (deg) &  & \\
\hline
    2017 &         1 & $307.50000$ & $-40.00000$ &       3 &      32 \\
    2017 &         2 & $308.96859$ & $-38.05144$ &       3 &      31 \\
    2017 &         3 & $307.50000$ & $-36.10289$ &       3 &      31 \\
    2017 &         4 & $308.96859$ & $-34.15433$ &       3 &      31 \\
    2017 &         5 & $310.43715$ & $-40.00000$ &       3 &      31 \\
    2017 &         6 & $311.90575$ & $-38.05144$ &       3 &      30 \\
    2017 &         7 & $310.43715$ & $-36.10289$ &       3 &      30 \\
    2017 &         8 & $311.90575$ & $-34.15433$ &       3 &      30 \\
    2017 &         9 & $313.37434$ & $-40.00000$ &       3 &      29 \\
    2017 &        10 & $314.84293$ & $-38.05144$ &       3 &      29 \\
    2017 &        11 & $313.37434$ & $-36.10289$ &       3 &      29 \\
    2017 &        12 & $314.84293$ & $-34.15433$ &       3 &      27 \\
    2017 &        13 & $316.31149$ & $-40.00000$ &       3 &      29 \\
    2017 &        14 & $317.78008$ & $-38.05144$ &       3 &      29 \\
    2017 &        15 & $316.31149$ & $-36.10289$ &       3 &      29 \\
    2017 &        16 & $317.78008$ & $-34.15433$ &       3 &      29 \\
   2018A &         1 & $172.50000$ & $-33.00000$ &       4 &      24 \\
   2018A &         2 & $173.84140$ & $-34.94856$ &       4 &      24 \\
   2018A &         3 & $172.50000$ & $-36.89711$ &       3 &      24 \\
   2018A &         4 & $175.18282$ & $-33.00000$ &       3 &      24 \\
\hline
\end{tabular} 

\end{subtable}
\begin{subtable}{.52\linewidth}
\centering
\begin{tabular}{|c|c|c|c|c|c|}

\hline

Campaign   & Field ID & R.A. & Dec. & N$_g$ & N$_r$ \\

&  & (deg) & (deg) &  & \\
\hline
   2018A &         5 & $176.52423$ & $-34.94856$ &       3 &      24 \\
   2018A &         6 & $175.18282$ & $-36.89711$ &       3 &      24 \\
   2018A &         7 & $177.86563$ & $-33.00000$ &       3 &      24 \\
   2018A &         8 & $179.20704$ & $-34.94856$ &       3 &      24 \\
   2018A &         9 & $177.86563$ & $-36.89711$ &       3 &      24 \\
   2018A &        10 & $180.54845$ & $-33.00000$ &       3 &      24 \\
   2018A &        11 & $181.88986$ & $-34.94856$ &       3 &      24 \\
   2018A &        12 & $180.54845$ & $-36.89711$ &       3 &      23 \\
   2018B &        13 & $232.50000$ & $-32.00000$ &       2 &      20 \\
   2018B &        14 & $233.82657$ & $-33.94856$ &       2 &      20 \\
   2018B &        15 & $232.50000$ & $-35.89711$ &       2 &      20 \\
   2018B &        16 & $235.15315$ & $-32.00000$ &       2 &      20 \\
   2018B &        17 & $236.47973$ & $-33.94856$ &       2 &      20 \\
   2018B &        18 & $235.15315$ & $-35.89711$ &       2 &      20 \\
   2018B &        19 & $237.80630$ & $-32.00000$ &       2 &      20 \\
   2018B &        20 & $239.13288$ & $-33.94856$ &       2 &      20 \\
   2018B &        21 & $237.80630$ & $-35.89711$ &       2 &      20 \\
   2018B &        22 & $240.45945$ & $-32.00000$ &       2 &      19 \\
   2018B &        23 & $241.78602$ & $-33.94856$ &       2 &      19 \\
   2018B &        24 & $240.45945$ & $-35.89711$ &       2 &      19 \\

\hline

\end{tabular}

\end{subtable}

\end{center}
\end{table*}

\begin{figure*}
\begin{center}
\includegraphics[angle=0,scale=.45]{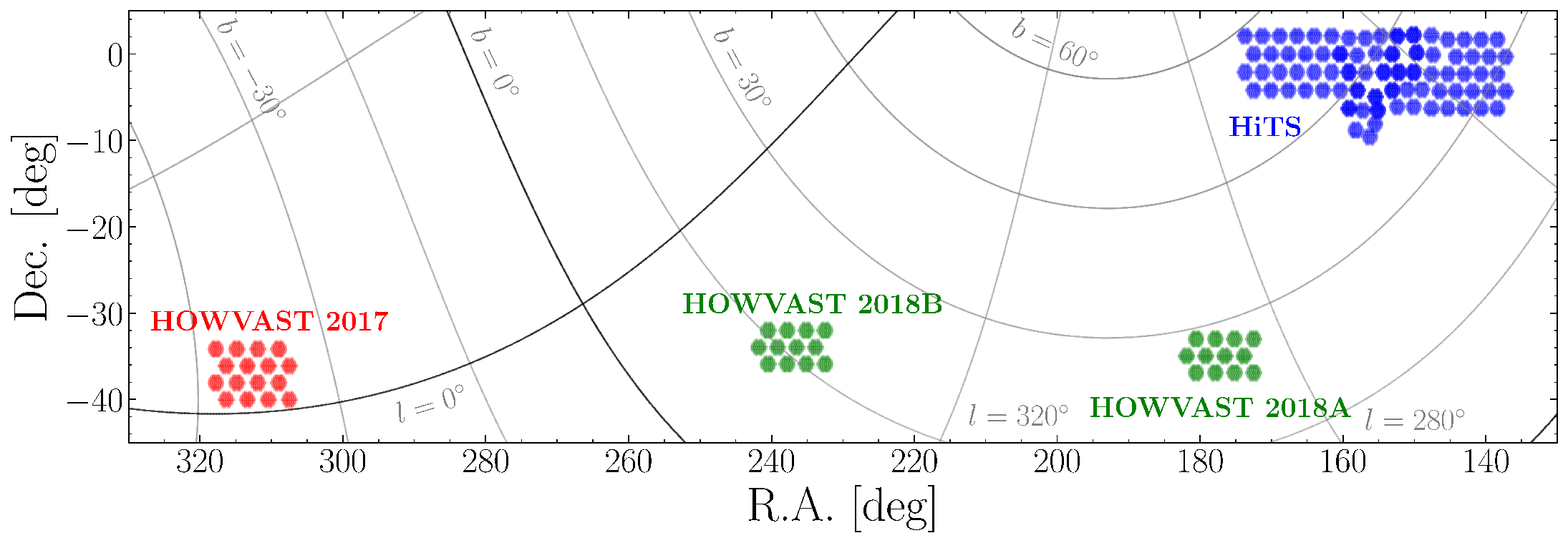}
\caption{
Spatial distribution of the surveys used for this work, shown in both equatorial and Galactic coordinates.
The DECam fields corresponding to the HOWVAST survey are plotted in red and green, while the fields observed by the HiTS survey are shown in blue.
}
\label{fig:skymap}
\end{center}
\end{figure*}

\section{The data}
\label{sec:data}

\subsection{Observations}
\label{sec:obs}

\subsubsection{DECam data}
\label{sec:DECam}

The data used in this work were obtained as part of three independent campaigns carried out with the Dark Energy Camera \citep[DECam;][]{Flaugher15}, which is mounted on the 4\,m telescope at Cerro Tololo Inter-American Observatory (CTIO). 
The first campaign corresponds to the HiTS survey, which was originally designed to characterize the early stages of supernovae explosions in real time \citep[][]{Forster16}.
Specifically, we use the data from HiTS that were observed between 2015 February 17 and 22 (program 2015A-0608, PI: F{\"o}rster). 
This region of the HiTS survey consists of 50 Galactic halo fields ($\sim$150\,deg$^2$) and includes 14 fields that were observed in previous HiTS campaigns (see Figure~\ref{fig:skymap}). 
The HiTS 2015 fields were observed up to five times per night, and are located between 137 and 160\,deg in right ascension, and $-7$ and $2.6$\,deg in declination \citep{Forster16}. 
The data were taken mainly in the $g-$band, with 87\,s exposures and a cadence of 1.6\,hr. 
Observations in the $r-$band were performed as well, with individual exposures ranging from 81 to 102\,s, which allowed the inclusion of $g-r$ colours for our analysis.
This configuration summed up a total of 20 to 29 epochs in $g$, and from one to ten in $r$, per field. 
For a more detailed description of HiTS' design, its observing strategy, and a comprehensive review of its characteristics we refer the reader to the work by \citet{Forster16}. 
It is worth mentioning that these data were not included in the work of \citet{Medina2018}, where a previous HiTS campaign (from year 2014) was analyzed.

The second and third observing campaigns took place on half-nights between 2017 August 26 and 31 (programs 2017B-0907, PI: Mu\~noz, and 2017B-0253, PI: Carlin) and full nights between 2018 April 20 and 23 (2018A-0215, PI: Carlin, and 2018A-0907, PI: Mu\~noz), in the context of the HOWVAST survey \citep[][]{Medina2021b}. 
For HOWVAST we selected DECam fields to cover a considerable range of Galactic latitudes of the MW halo (from $-43.0$ to $28.7$\,deg). 
The footprint of HOWVAST was chosen to avoid a significant overlap with deep large-sky surveys, such as the DES \citep[][]{DES2005,DES2016}. 
In the first HOWVAST observing run, we observed 16 DECam fields during four consecutive half-nights (in first quarter moon phase), with 180\,s exposures in the $r-$band and a cadence of approximately one hour.
The second HOWVAST campaign consists of 24 fields,  separated into two groups of 12 fields at different Galactic latitudes, observed in the $r-$band during four consecutive nights (in first quarter moon phase). 
As in the first run, the integration times for this campaign were of 180\,s, but with a cadence of $\sim$40\,min.
This results in a combined area of $\sim$120\,deg$^2$ mapped in the halo surveyed by HOWVAST, and time series containing from 15 to 30 epochs per star.
In addition, we obtained from two to four 240\,s-exposure epochs per field in the $g-$band each year in order to facilitate, thanks to the colour information,  the identification of RRLs in our analysis.
The coordinates of the HOWVAST fields in the equatorial system (J2000), including the number of observations per field, are provided in Table~\ref{tab:fields}, making the distinction between the high- and low-Galactic latitude fields observed during our second campaign. 

Combining the footprints of HiTS and HOWVAST, in this work we analysed an area of $\sim$270\,deg$^2$ to search for RRLs. 
To increase the sample size of halo RRLs for our posterior analysis, we complement our detections with the RRL catalogue of \citet{Medina2018} (which covers $\sim$80\,deg$^2$ of additional area). 
Therefore, in this work we analysed a total of $\sim$350\,deg$^2$. 
The sky coverage of the entire HiTS campaign and HOWVAST is shown in Figure~\ref{fig:skymap} in equatorial and Galactic coordinates. 

\subsection{Data processing}
\label{sec:dataProc}

\subsubsection{Pre-processing}
\label{sec:dataPreProc}

The data reduction for this work was performed using the DECam community pipeline \citep{Valdes2014}. 

The data from the HiTS 2015 campaign were pre-processed as part of the work by \citet{Medina2017thesis}. 
A catalogue with the sources in these fields was created following the methodology from \citet{Medina2018}. 
We first defined an x, y pixel coordinate system based on the output information generated by the SExtractor photometry software \citep{Bert96}. 
To do this, we selected a reference frame for both the $g$ and the $r-$band, for which the observing conditions were closer to optimal (i.e., from a photometric night and with the best seeing).  
Subsequently, we used the scaling constants found by the HiTS pipeline \citep{Forster16} to perform the alignment of the individual observations with respect to the reference.
Then, we cross-matched the catalogues aligned in the common x, y coordinate system and rejected sources with fewer than five detections in the $g-$band for the rest of the analysis.
In order to keep sources with preliminary indications of variability for further processing, we disregarded sources for which the uncertainties in the mean flux exceeded by more than two times their flux standard deviations.
Finally, we applied the x, y pixel to equatorial coordinate transformation using the astrometric solutions derived by \citet[][]{Forster16}, which is computed from the crossmatch with known stars in the USNO catalog \citep[][]{Monet2003}.

To pre-process the HOWVAST observations we adopted an alternative approach, based on the data processing pipeline in development for the Rubin Observatory Legacy Survey of Space and Time \citep[LSST;][]{LSST2009,Bosch2019}. 
This pipeline was used to detect sources in the images, measure aperture fluxes, and perform a source point spread function (PSF) fitting.
Because the sources of interest of this work are stars, we use PSF fluxes and magnitudes throughout the HOWVAST data treatment. 
For the subsequent variable star analysis, we only examined stellar sources with more than 15 data points in their time series, and with flux stellar deviations at least 2.5 times larger than their mean flux uncertainties.

\vspace{1.0cm}

\subsubsection{Photometric calibration}
\label{sec:photCal}

In order to account for atmospheric effects affecting the epochs in our time series, we determined a photometric zero-point relative to the reference frame chosen for HiTS and HOWVAST separately.
For this, we first computed instrumental magnitudes following

\begin{equation}
{\rm mag}_{\rm inst} = -2.5\, \log \left(\frac{\rm Flux}{t_{\rm exp}}\right) - a_g - k_g\, A
\label{eq:maginst}
\end{equation}

\noindent where ${\rm mag}_{\rm inst}$ represents the instrumental magnitude either in the $g$ or the $r$ filter, ${\rm Flux}$ is the source flux, $t_{\rm exp}$ corresponds to the exposure time, $a$ and $k$ are the filter-dependent DECam photometric zero-point and first-order extinction coefficient per CCD\footnote{\scalebox{0.8}{Available at \url{www.noirlab.edu/science/documents/scidoc1571}}  } (respectively), and $A$ is the airmass at the time of the observations.  

To calibrate the photometry of the HiTS 2015 fields, we first anchored our instrumental magnitudes to the reference frames in $g$ and $r$. 
For the photometric calibration of the HOWVAST 2017 and 2018 data, we selected reference frames in which the PSF of the sources were minimum, similar to what was done for the HiTS 2015 data pre-processing.
We compared our instrumental magnitudes with those in the reference frame, so that ${\textnormal{\rm mag}}_{\textnormal{\rm ref}} = {\textnormal{\rm mag}}_{\textnormal{\rm inst}} + \Delta_{\rm rel}$, where $\Delta_{\rm rel}$ is the zero-point relative to the reference epoch, and ${\textnormal{\rm mag}}_{\textnormal{\rm ref}}$ is the object magnitude calibrated with respect to said epoch.

We then calibrated the photometry of the references using the archival data stored in the National Optical-Infrared Astronomy Research Laboratory (NOIRLab) Source Catalog \citep[NSC;][]{Nidever2021},  
as all the surveys considered in our work overlap the catalogs published in the second data release of such database. 
The photometric calibration of the NSC is based on the PS-1 survey, on the Skymapper 
and the Asteroid Terrestrial-impact Last Alert System (ATLAS) all-sky stellar reference catalogues (\citealt{Wolf2018} and \citealt{Tonry2018}, respectively), 
and on model magnitudes from linear combinations of photometry from catalogs such as the Two Micron All-Sky Survey \citep[2MASS;][]{Skrutskie2006} and the American Association of Variable Star Observers (AAVSO) Photometric All-Sky Survey \citep[APASS;][]{Henden2015}. 
For the calibration, we limited the NSC data to star-like sources, with {\it starClass} flags larger than 0.85 (a {\it starClass} value of 0 is assigned for extended sources in the NSC, and a value of 1 is used for point-like sources).  
The sample was selected to include the best-matching NSC star within two arcseconds from each of the sources in our catalogue. 
We used only NSC stars with $g$ and $r$ magnitudes between 16.5 and 20.5, and magnitude errors smaller than 0.05, in addition to a two sigma clipping process performed over the median magnitude difference to remove outliers. 
From this comparison, we obtained an additional zero-point and colour term on a chip-by-chip basis, for each DECam field.
Therefore, the calibrated magnitudes are given by

\begin{equation}
\begin{array}{lc}
{\rm mag} = {\rm mag}_{\rm ref} - A_{\rm NSC} - B_{\rm NSC} \cdot (g - r)
\end{array}
\label{eq:mags}
\end{equation}

\noindent where mag represents the calibrated magnitudes, $A_{\rm NSC}$ and $B_{\rm NSC}$ are the zero-point and the colour coefficient resulting from the magnitude comparison with the NSC data, respectively, and $g - r$ is the colour of a given star. 
The average root mean square of these calibrations (across all fields and CCDs) 
is approximately 0.02\,mag for the HiTS and the 2018 fields (in both bands), and of 0.03\,mag for the 2017 fields. 
We note that these values are sufficiently small for the purpose of our work, but that calibrating our magnitudes onto {\it Gaia} synthetic photometry \citep[see][]{GaiaMontegriffo2023} could contribute to an overall decrease if needed. 

Finally, the mean magnitudes were 
corrected for reddening using the dust maps of \citet{SFD98}, adopting the extinction coefficients (i.e., $A_g = 3.237\ E(B-V)$ and $A_r = 2.176\ E(B-V)$) from \citet{Sch11}. 
Magnitude errors were computed by propagation of uncertainties. 

\begin{figure}
\begin{center}
\includegraphics[angle=0,scale=.55]{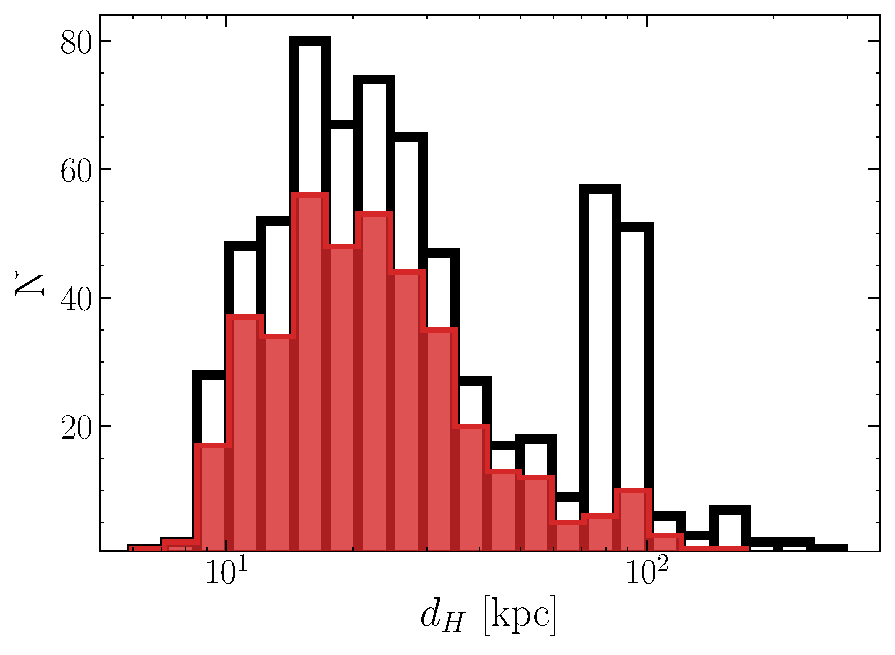}
\caption{
Distribution of the heliocentric distances $d_{\rm H}$ of the RRLs analysed in our work. 
The red-filled region represents the distance of the stars detected in this work (as described in Section~\ref{sec:distances}), whereas black bars depict the distribution of the full RRL sample, including those from \citet{Medina2018}. }
\label{fig:histogram}
\end{center}
\end{figure}

\begin{figure*}
\begin{center}
\includegraphics[angle=0,scale=.50]{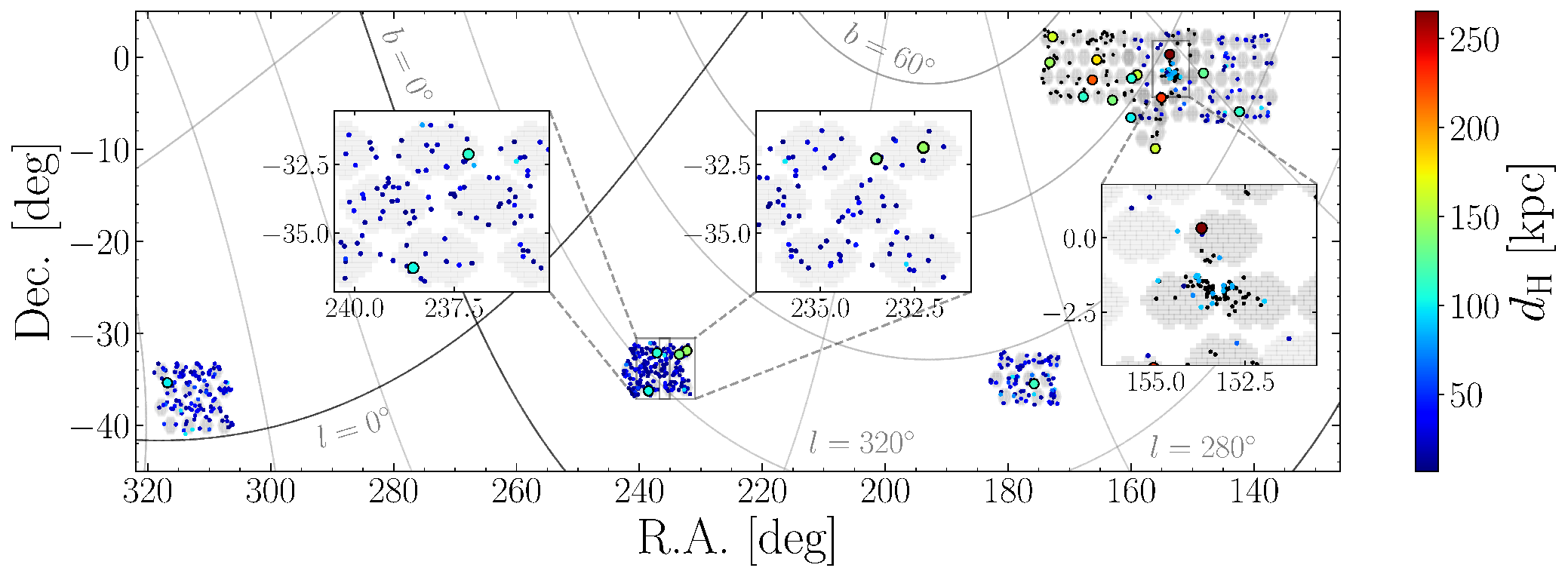}
\caption{
Spatial distribution of the RRLs analysed in this work, colour-coded by their heliocentric distance $d_{\rm H}$ in kpc. 
The stars from the HiTS campaign analyzed by \citet{Medina2018} are plotted with small black symbols, and the RRLs from this work with distances larger than 100\,kpc (described in Section~\ref{sec:100kpc}) are plotted with large colour-coded circles. 
An approximation of the footprint of each DECam field observed is shown in gray in the background as a reference.
Enlargements of the two regions containing RRLs beyond 100\,kpc with similar distances (and potentially associated with each other) are provided as insets. 
A third enlarged region depicts the overdensity associated with the Sextans dSph. 
}
\label{fig:skymapRRLs}
\end{center}
\end{figure*}

\begin{table*}\small
\caption{
Total number ($N$) and recovered RRLs ($N_x$) from previous surveys as a function of heliocentric distance $d_{\rm H}$. This table shows that, while our survey is partially limited to sources with mean magnitudes fainter than 17.5 in $g$ and $r$, we are able to recover over 70\,per\,cent of the RRLs beyond 20\,kpc. 
}
\label{tab:completi}
\begin{center}

\begin{tabular}{|c|HHHHccc|HHHccc|ccc|ccc|}
\hline

\multicolumn{1}{|c|}{Survey} &
\multicolumn{1}{H}{Survey (ours)} &
\multicolumn{3}{H}{All} &
\multicolumn{3}{c|}{All $d_{\rm H}>$ 20\,kpc} &
\multicolumn{3}{H}{$d_{\rm H}<$ 20\,kpc} &
\multicolumn{3}{c|}{20 $<d_{\rm H}<$ 40\,kpc} &
\multicolumn{3}{c|}{40 $<d_{\rm H}<$ 80\,kpc} &
\multicolumn{3}{c|}{$d_{\rm H}>$ 80\,kpc} \\
\hline
 & & $N$ & $N_x$ & $N_x$/$N$ & $N$  & $N_x$ & $N_x$/$N$ & $N$  & $N_x$ & $N_x$/$N$  & $N$ & $N_x$ & $N_x$/$N$  & $N$ & $N_x$  & $N_x$/$N$  & $N$ & $N_x$ & $N_x$/$N$ \\
\hline
   GaiaDR3 &  HiTS 2015 + HV &          896 &         392 &            0.44 &                 292 &                179 &                   0.61 &               494 &              188 &                 0.38 &                   173 &                  125 &                     0.72 &                    75 &                   31 &                     0.41 &                  44 &                 23 &                   0.52 \\
  PS-1 &  HiTS 2015 + HV &          467 &         147 &            0.31 &                 104 &                 58 &                   0.56 &                67 &               30 &                 0.45 &                    32 &                   22 &                     0.69 &                    31 &                   17 &                     0.55 &                  41 &                 19 &                   0.46 \\
        CS &  HiTS 2015 + HV &          554 &         239 &            0.43 &                 152 &                107 &                   0.70 &               341 &              120 &                 0.35 &                   137 &                   97 &                     0.71 &                    15 &                   10 &                     0.67 &                   0 &                0 &                     -- \\
       ZTF &  HiTS 2015 + HV &           87 &          37 &            0.43 &                  13 &                  7 &                   0.54 &                31 &                9 &                 0.29 &                    11 &                    6 &                     0.55 &                     2 &                    1 &                     0.50 &                   0 &                0 &                     -- \\
       DES &  HiTS 2015 + HV &            5 &           2 &            0.40 &                   3 &                  2 &                   0.67 &                 2 &                0 &                 0 &                     0 &                  0 &                       -- &                     2 &                    2 &                     1.00 &                   1 &                  0 &                   0.00 \\
\hline
\end{tabular}
\end{center}
\end{table*}

\begin{figure}
\begin{center}
\includegraphics[angle=0,scale=0.35]{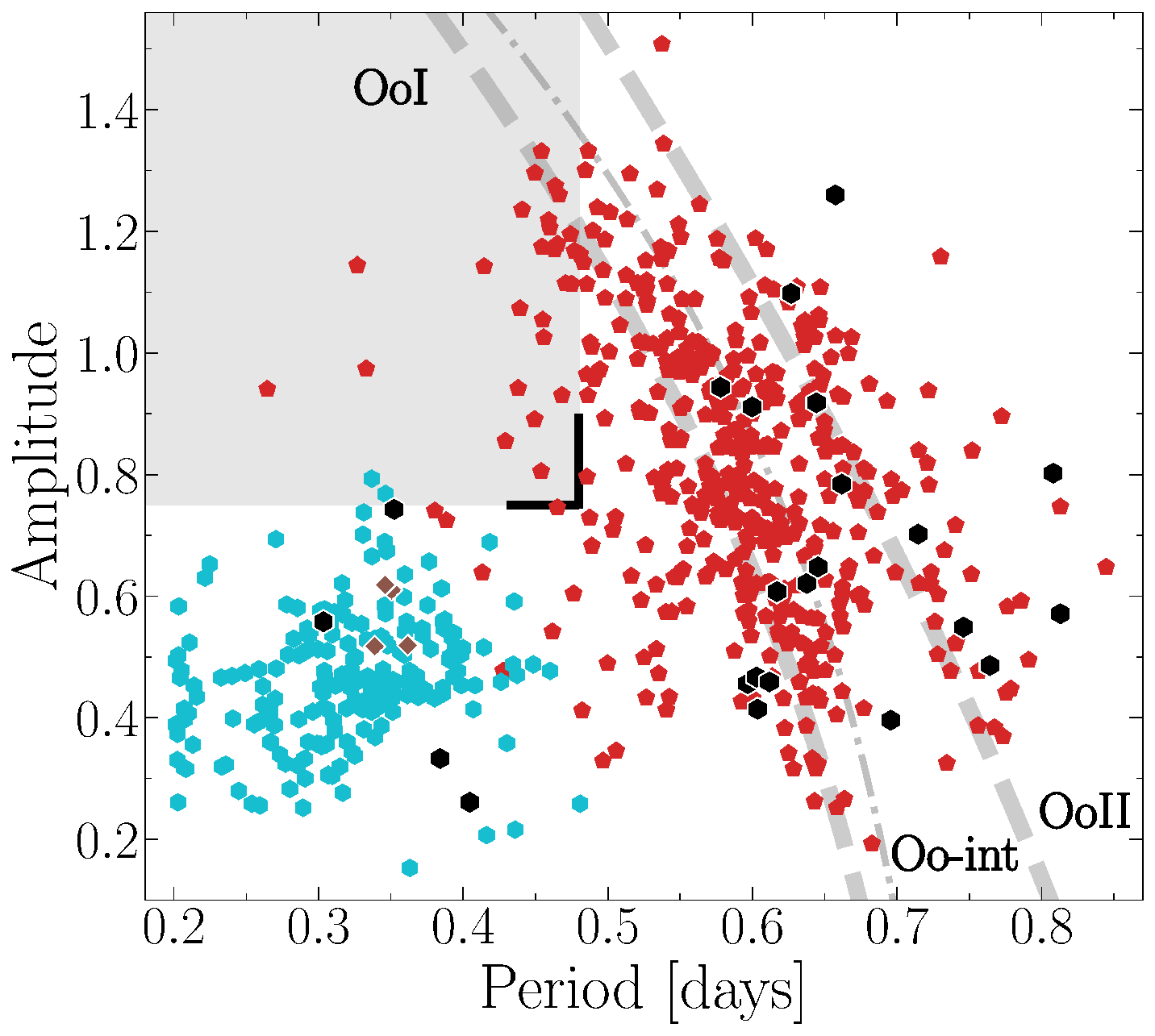}
\caption{
Bailey diagram of the RRLs analysed in this work. 
RRab stars are displayed with red pentagons, whereas light-blue and brown 
symbols represent c-type and d-type RRLs, respectively. 
The amplitudes represent the minimum-to-maximum variation of the fitted light curves in the $V$ band, obtained by scaling the $g$ and $r$ magnitude amplitudes by a factor of 0.90 and 1.21, respectively \citep{Sesar2012}.
Stars with estimated heliocentric distances larger than 100\,kpc are plotted with black symbols.
The dashed regions depict the fiducial lines for OoI, Oo-int, and OoII defined by \citet{Fabrizio2019}.
Black solid lines delimit the region containing HASP RRab variables (as defined by \citealt{Fiorentino2015}), which is shown as a grey shaded area. 
}
\label{fig:bailey}
\end{center}
\end{figure}

\section{Search and Characterization of RR Lyrae}
\label{sec:search}

\subsection{Selection of the RR Lyrae} 
\label{sec:selection}

Since the data used in this work were obtained from two different surveys, we note that two slightly different methodologies were adopted to process the data for the search and characterization of RRLs. 

The first sample corresponds to the RRLs identified in the HiTS 2015 survey by \citet{Medina2017}. 
When looking for RRL candidates in this survey, objects with a magnitude variation smaller than 0.2 magnitudes were filtered out from the original source catalogues. 
Although applying this cut undoubtedly hinders the detection of low-amplitude RRLs, we consider this necessary in order to avoid contamination from non-variables in the faint end of the survey, where photometric uncertainties can easily reach 0.1\,mag. 
Additionally, only sources redder than $-0.2$ in $g-r$, and bluer than $0.6$ were considered for further processing. 
These values were chosen to remove potential spurious RRLs falling outside of the instability strip and to reduce the number of candidates to be inspected \citep[as done by, e.g.,][]{Vivas2019}. 

\vspace{0.5cm}

The period of the sources in the HiTS 2015 catalogues were determined by \citet{Medina2017} based on the generalized version of the Lomb-Scargle period detection routine \citep[GLS;][]{Zech09}, which incorporates a constant in the typical Lomb-Scargle sinusoid fitting procedure.
By doing this, the results are overall less susceptible to aliasing and provide a more accurate frequency selection in the power spectrum.
To compute the statistics and period selection, the GLS tool within the \textit{astroML} Python module \citep{Vander12} was used.
Only sources with periods longer than $4.8$\,hours ($0.2$\,d) and shorter than $21.6$\,hours ($0.9$\,d), as well as those for which the GLS statistical level detections were smaller than $0.08$, were considered in order to reduce the number of RRL candidates.  
We note in passing that these cuts should encompass most of the periods observed in RRLs \citep[][]{Catelan2015} 
while avoiding aliasing around the period of one day and removing contamination from short period BL Her stars. 
For comparison, none of the RRL in the {\it Gaia} DR3 catalogue \citep[][]{Clementini2022} have periods shorter than 0.2\,d, and only 0.13\,per\,cent of them display periods longer than 0.9\,d (0.05\,per\,cent longer than 0.95\,d). 
Finally, the two most significant periods were chosen and further inspected when more than one period were detected and met the aforementioned requirements. 
The last step for the selection of RRLs in the HiTS 2015 fields was to visually inspect the light curves resulting from the previous cuts and to look for objects with light-curve shapes, periods, and amplitudes typical of RRLs.
The search resulted in a total of 95 RRLs in the fields that do not overlap with those inspected by \citet{Medina2018}.

The selection criteria for RRLs in HOWVAST data are similar to the one followed for HiTS. 
We filtered out sources with minimum to maximum magnitude variability smaller than 0.2\,magnitudes, and those with $g-r$ colours clearly differing from the expectations for RRLs.
In this case, given that a subset of the HOWVAST fields lie in regions with non-negligible interstellar extinction (at lower Galactic latitudes), 
we adopted a more generous cut and considered stars with $g-r$ between $-0.25$ and $1.0$ for further analysis.

\begin{table*}
\scriptsize
\caption{
Main properties of the sample of distant RRLs discussed in Section~\ref{sec:100kpc}, including equatorial coordinates, mean magnitude and number of epochs per filter, and heliocentric and Galactocentric distances. 
We flag the stars detected in this work with zeros and those from \citet{Medina2018} with ones. 
}
\label{tab:distant}
\begin{center}
\begin{tabular}{|c|c|cH|c|c|c|c|c|c|c|c|c|c|c|}

\hline
ID &           R.A. &          DEC. &       From &           $<g>$ &           $<r>$ & $N_g$ & $N_r$ &    Period$^*$ & Amplitude$^*$ & Type & $d_{\rm H}$ & $R_{\rm sph}$ & $R_{\rm el}$ & Flag \\
  & (deg) & (deg) & &   &  & & & (days ) &  & & (kpc) & (kpc) & (kpc)   &  \\
\hline

   HV210719-352250 &  $316.82756$ &  $-35.38046$ &  decam2017 &  $20.79\pm0.03$ &  $20.68\pm0.01$ &   $3$ &  $28$ &  $0.6028$ &    $0.39$ &   ab &   $100\pm3$ &     $95\pm3$ &    $117\pm4$ &  $0$ \\
 HiTS104009-063304 &  $160.03895$ &   $-6.55105$ &   HiTS2014 &  $20.66\pm0.01$ &              -- &  $21$ &    -- &  $0.6376$ &    $0.69$ &   ab &   $100\pm4$ &    $102\pm4$ &    $125\pm5$ &  $1$ \\
   HV155407-361645 &  $238.52769$ &  $-36.27926$ &  decam2018 &  $22.01\pm0.03$ &  $21.31\pm0.01$ &   $2$ &  $19$ &  $0.6618$ &    $0.65$ &   ab &   $104\pm4$ &     $96\pm3$ &     $99\pm4$ &  $0$ \\
 HiTS103943-021726 &  $159.93119$ &   $-2.29061$ &   HiTS2014 &  $20.74\pm0.01$ &              -- &  $21$ &    -- &  $0.6956$ &    $0.44$ &   ab &   $108\pm4$ &    $110\pm4$ &    $136\pm5$ &  $1$ \\
   HV154834-320810 &  $237.14125$ &  $-32.13607$ &  decam2018 &  $21.57\pm0.02$ &  $20.94\pm0.01$ &   $2$ &  $19$ &  $0.5967$ &    $0.38$ &   ab &   $110\pm4$ &    $103\pm3$ &    $108\pm4$ &  $0$ \\
 HiTS111106-041718 &  $167.77512$ &   $-4.28834$ &   HiTS2014 &  $20.99\pm0.01$ &              -- &  $22$ &    -- &  $0.3034$ &    $0.62$ &    c &   $111\pm4$ &    $113\pm4$ &    $143\pm6$ &  $1$ \\
 HiTS092927-055440 &  $142.36180$ &   $-5.91120$ &   HiTS2015 &  $21.09\pm0.01$ &  $20.98\pm0.01$ &  $25$ &    -- &  $0.5779$ &    $1.05$ &   ab &   $112\pm4$ &    $116\pm4$ &    $131\pm5$ &  $0$ \\
   HV114307-352948 &  $175.78118$ &  $-35.49667$ &  decam2018 &  $21.27\pm0.02$ &  $21.01\pm0.01$ &   $3$ &  $24$ &  $0.5998$ &    $0.75$ &   ab &   $115\pm4$ &    $113\pm4$ &    $124\pm4$ &  $0$ \\
 HiTS095253-014305 &  $148.22270$ &   $-1.71800$ &   HiTS2015 &  $21.35\pm0.01$ &  $21.48\pm0.02$ &  $27$ &    -- &  $0.3523$ &    $0.83$ &    c &   $128\pm5$ &    $132\pm5$ &    $155\pm6$ &  $0$ \\
   HV153403-321831 &  $233.51328$ &  $-32.30856$ &  decam2018 &  $21.45\pm0.02$ &  $21.31\pm0.01$ &   $2$ &  $18$ &  $0.8080$ &    $0.66$ &   ab &   $135\pm5$ &    $128\pm4$ &    $136\pm5$ &  $0$ \\
 HiTS105209-043942 &  $163.03718$ &   $-4.66174$ &   HiTS2014 &  $21.35\pm0.01$ &              -- &  $20$ &    -- &  $0.6036$ &    $0.46$ &   ab &   $136\pm5$ &    $138\pm5$ &    $171\pm6$ &  $1$ \\
   HV152905-315335 &  $232.27196$ &  $-31.89316$ &  decam2018 &  $21.97\pm0.03$ &  $21.50\pm0.01$ &   $2$ &  $16$ &  $0.8129$ &    $0.47$ &   ab &   $144\pm5$ &    $137\pm5$ &    $146\pm5$ &  $0$ \\
 HiTS103601-015451 &  $159.00456$ &   $-1.91422$ &   HiTS2014 &  $21.54\pm0.02$ &              -- &  $21$ &    -- &  $0.4046$ &    $0.29$ &    c &   $159\pm5$ &    $161\pm6$ &    $199\pm7$ &  $1$ \\
 HiTS102414-095518 &  $156.05905$ &   $-9.92180$ &   HiTS2014 &  $21.55\pm0.02$ &              -- &  $21$ &    -- &  $0.7641$ &    $0.54$ &   ab &   $161\pm7$ &    $163\pm7$ &    $192\pm8$ &  $1$ \\
 HiTS110222-001624 &  $165.59251$ &   $-0.27337$ &   HiTS2014 &  $21.95\pm0.02$ &              -- &  $19$ &    -- &  $0.6118$ &    $0.51$ &   ab &   $180\pm7$ &    $181\pm7$ &    $232\pm9$ &  $1$ \\
 HiTS110510-022710 &  $166.28982$ &   $-2.45282$ &   HiTS2014 &  $22.23\pm0.02$ &              -- &  $19$ &    -- &  $0.7459$ &    $0.61$ &   ab &   $219\pm9$ &    $220\pm9$ &   $280\pm11$ &  $1$ \\
 HiTS102014-042354 &  $155.05789$ &   $-4.39843$ &   HiTS2014 &  $22.37\pm0.03$ &              -- &  $19$ &    -- &  $0.3841$ &    $0.37$ &    c &   $229\pm8$ &    $231\pm8$ &   $279\pm10$ &  $1$ \\
 HiTS101453+001915 &  $153.71970$ &    $0.32090$ &   HiTS2015 &  $22.88\pm0.03$ &  $22.39\pm0.05$ &  $22$ &    -- &  $0.6169$ &    $0.67$ &   ab &  $265\pm11$ &   $268\pm11$ &   $327\pm13$ &  $0$ \\

\hline

\end{tabular} 
 \vspace{0.001cm}
 
     $^*$The period and amplitude of pulsation are computed from the photometric band with more observations.\\
\end{center}

\end{table*}

To determine periods for the remaining sources, we used the Python package P4J\footnote{\scalebox{0.8}{Available at \url{https://www.github.com/phuijse/P4J}.}}, which was specifically designed for period detection on irregularly sampled and heteroscedastic time series, using the Cauchy-Schwarz Quadratic Mutual Information (QMI) as the criterion to be maximized by this routine \citep{Huijse18}.
We first inspected the two periods with highest likelihoods, as long as they were longer than 0.2\,d and shorter than 0.90\,d and considered to have high statistical significance (at a 0.01 level). 
It should be noted that adopting these filters makes the most significant selection filter for HOWVAST's RRLs, namely the period detection criterion, comparable with the one used for HiTS' candidates. 
In fact, applying the QMI method to the 
RRL sample of HiTS yields periods that are indistinguishable from the GLS-based ones, with a median difference of the order of $10^{-5}$\,d. 
This similarity is not an unexpected result, as \citet{Huijse18} argue that the difference in the ability to recover the period of an RRL
between both methods is minimized for time series containing over 20 epochs. 
We refer the reader to \citet{Huijse18} for a detailed description of the similarities and differences of both methods. 

Finally, we visually inspected the phased light curves and selected only RRL-like objects, based on their light curve shapes, periods, and amplitudes. 
For stars exhibiting more than two high probability signals in the power spectrum, we examined the four most likely periods before choosing the star's main period.  

The final list of RRLs from the HOWVAST data consists of 397 stars. 
Thus, by considering both RRLs from HiTS and HOWVAST, in this work we report the detection of a total of 
492 RRLs. 
Their main properties are provided in Table~\ref{tab:fulltable}.  

In the remainder of this work, and for the sake of improving the number statistics of our analysis, we complement our sample with the RRLs from \citet{Medina2018}. 
Hereafter, we refer to this updated catalogue as the {\it full} or {\it combined} sample. 
With this, we increase our catalogue size from 492 to a total of 663 RRLs.

\subsection{Distance determination}
\label{sec:distances}

We determined the absolute magnitude of our RRLs in the $g$ and $r$ bands ($M_g$ and $M_r$, respectively) using the period-luminosity-metallicity relations for PS1 filters from \citet{Sesar2017} and assuming halo metallicity ([Fe/H]$ =-1.5$, close to the peak of the halo metallicity distribution; see e.g. \citealt{Suntzeff1991,Prantzos2008,Conroy2019}):

\begin{equation}
\begin{array}{lc}
M_g = \left(-1.7\pm 0.3\right)\, \log \left( \frac{P}{0.6} \right) + \left(0.69\pm 0.08\right)\\
M_r = \left(-1.6\pm 0.1\right)\, \log \left( \frac{P}{0.6} \right) + \left(0.51\pm 0.07\right)
\end{array}
\label{eq:PLZs}
\end{equation}

\noindent where $P$ stands for the periods of the RRLs, and the second term takes into account the uncertainty of the zero point of the relations and their metallicity dependence. 
Given that these relations are only valid for fundamental-mode periods, for RRc stars we ``fundamentalize'' their periods prior to using Equation~\ref{eq:PLZs} by following:

\begin{equation}
\log \left( P_{\rm F} \right) = \log \left( P \right) + 0.128
\label{eq:PFs}
\end{equation}

\noindent where $P_{\rm F}$ is the fundamentalized period \citep{Catelan2009}.
Heliocentric distances $d_{\rm H}$ are then computed through the distance modulus, and their  uncertainties determined from error propagation.
We note that computing metallicities from the Fourier decomposition of our light curves yields a median [Fe/H] of $-1.4$\,dex (with individual uncertainties of $\sim$0.3\,dex) when using the period–$\phi_{31}$–[Fe/H] relation derived by \citet{Mullen2021}. 
This approach, however, is not directly applicable to all of our RRLs due to the sparsely sampled light curves of a significant fraction of the sample. 
The effects of our metallicity assumption on the resulting distances in Equation~\ref{eq:PLZs} are expected to be small. 
In fact, an offset of 0.5 and 1.0\,dex in [Fe/H] would lead to differences in $d_{\rm H}$ smaller than 4 and 8\,kpc for remote RRLs ($> 100$\,kpc), respectively. 

Figure~\ref{fig:histogram} displays the heliocentric distance distribution of our RRLs and of those from the combined sample (i.e., considering the RRLs from \citealt{Medina2018}). 
Our sample consists of RRLs with $d_{\rm H}$ spanning from $7$ to $\sim$270\,kpc.
Most of these stars lie within 50\,kpc (429 RRLs; 87.2\,per\,cent), whereas 54 of them have $d_{\rm H}$ between 50 and 100\,kpc (10.9\,per\,cent), and 11 (1.8\,per\,cent) lie beyond 100\,kpc. 
We further describe the most distant subsample in Section~\ref{sec:100kpc}.
An overdensity of 16 RRLs near 85\,kpc ($< g > \sim $ 20.5 or ($< r > \sim $ 20.2) is clear from the figure, and is associated with RRLs in the Sextans dwarf spheroidal galaxy (dSph) that were not detected by \citet{Medina2018}. 
We note that, similar to \citet{Medina2018}, we find a discrepancy of $\sim10$\,kpc between the mean distance to Sextans from RRab- and RRc-only samples ($\sim84$\,kpc and $\sim75$\,kpc, respectively). 
This offset (of $11\pm6$\,per\,cent) may be caused by the fact that the period-luminosity-metallicity relations from \citet{Sesar2017} are derived for fundamental-mode RRLs only or by differences in the passbands used in our works.  
Thus, we corrected the distances to the entire RRc stars in our sample (including those shown in Figure~\ref{fig:histogram}) by this factor.

\citet{Vivas2019} searched for periodic variables in the Sextans dSph and all of our stars are found in their catalog when crossmatching within a radius of 1.5\,arcsec, with the exception of the stars HiTS100752-020827 ($d_{\rm H}\sim80.5$\,kpc), HiTS101128-013921 ($d_{\rm H}\sim82.8$\,kpc), HiTS101338-015258 ($d_{\rm H}\sim80.1$\,kpc), and HiTS101734+001322 ($d_{\rm H}\sim86.9$\,kpc), which are new detections. 
We highlight that our RRL classification matches with that of \citet{Vivas2019} for all the stars in common, and that the mean (absolute) period difference is of $0.008$\,d. 
Finally, we determined the period of three RRLs for which no period was reported by \citet{Vivas2019}, namely HiTS101456-022025 (RR178 in their work, with a period of 0.4890\,d), HiTS101342-021246 (RR142, 0.3999\,d), and HiTS101551-015619 (RR192, 0.3232\,d).

\subsection{Completeness and recovery rate}
\label{sec:completeness}

Understanding how effective our methodology is in detecting RRLs and its limitations is crucial for interpreting the results of our search and for our RRL spatial distribution analysis. 
To quantify the effects of photometric completeness on our RRL detection efficiency, we take advantage of the fact that our data have been included in an early version of Data Release 3 of the DECam Local Volume Exploration (DELVE) survey (for an overview of DELVE, see \citealt{Drlica-Wagner2021}). 
DELVE includes new data observed specifically for the survey, but also aggregates those new data with all existing DECam data in the public archive. 
Thus, our HOWVAST RRL data are included in DELVE DR3 processing, wherein all available data are coadded to create deep, stacked images. 
We estimate our survey's photometric completeness by comparing our single exposure catalogues with DELVE's coadded catalogues.  
For the latter, we consider point sources only (``high confidence stars,'' following the definition of \texttt{EXTENDED\_CLASS} from Section~4.7 of \citealt{DES_DR2_2021}), with DELVE photometric uncertainties $<0.5$\,mag and number of epochs greater than zero. 
Then, we crossmatch these coadded catalogs with ours and compute the fraction of recovered stars as a function of magnitude. 
Our survey recovers over 95\,per\,cent of the DELVE sources brighter than $r\sim22$\,mag consistently across epochs and the recovery fraction drops to $\sim50$\,per\,cent at $r\sim23$\,mag.
The photometric completeness curves obtained with this strategy are very similar to those of the HiTS survey presented by \citet{MartinezPalomera2018} (Figure 2 in their paper), which is not surprising given the similarity in observing conditions, strategies, and designs between both surveys.

The next step was to generate 5000 artificial RRL light curves mimicking the cadence and the expected photometric uncertainties (as a function of magnitude) of our survey. 
We then applied our photometric completeness (per epoch) estimates to the time series of a given light curve, based on the likelihood that an RRL would be observed at a given epoch (which is a function of magnitude as well). 
This typically results in a smaller number of epochs per light curve than in the ideal case (i.e., if the photometric completeness was 100\,per\,cent), especially for fainter magnitudes. 
We run our selection pipeline and label as recovered the stars that pass our filters (see Section~\ref{sec:selection}) and have a computed period within 10\,per\,cent of the real (i.e., simulated) one.  
We find that, in this ideal scenario, we expect up to 90-95 per cent of the real RRLs to be recovered within 100 kpc. These numbers decrease significantly for sources fainter than $r>21.5$ ($\sim150$\,kpc), where the recovery rate drops to $<80$\,per\,cent. 
However, since there are other effects in play for real observations of individual sources (e.g., blending of sources, proximity to bright sources, and their positions in the CCDs), the number of recovered RRLs should be slightly smaller overall than predicted by our idealized simulation.  

\vspace{1cm}

\subsection{Comparison with previous surveys}
\label{sec:comparison}

To assess the fraction of RRLs from other surveys that might be missing in our catalog we crossmatch our sample with the RRLs in the Catalina Real-time Transient Survey \citep[CRTS;][]{Dra13,Dra14,Drake17,Torrealba2015}, the PS-1 \citep[][]{Sesar2017} survey, and with the {\it Gaia} catalogue generated with the SOS pipeline. 
For the latter, we use the recently published catalogue based on {\it Gaia} DR3 \citep{Clementini2022}, which almost doubles the size of its predecessor, {\it Gaia} DR2 \citep{Clementini2019}.
We perform our crossmatches using a search radii of 3\,arcsec to account for the different pixel scales of these surveys. 
For these comparisons, we only considered the RRLs from the literature that fall within the footprint of our survey, excluding the DECam's CCD gap regions. 
We emphasize that, by design, HOWVAST does not significantly overlap the area covered by large surveys such as the DES \citep[][]{Stringer21} 
and the ZTF \citep[][]{Chen2020,Huang2022} survey. 
The results of our comparisons, including the few overlapping RRLs from the DES and the ZTF are summarized in Table~\ref{tab:completi}.

We restrict our comparison to a meaningful magnitude (distance) range to avoid using stars in the CRTS, PS-1, or {\it Gaia} that could have saturated epochs 
in our survey ($\lessapprox 20$\,kpc). 
When limiting the comparison to RRLs between 20 and 40\,kpc, observed at least 70\,pixels from the edge of the CCDs (and with detections in the $BP$ and $RP$ bands, for {\it Gaia}), the number of recovered RRLs is close 70-75\,per\,cent in each case with the exception of the ZTF, where we recover $\sim$55\,per\,cent of the RRLs.  
If we expand the distance range to the region spanning from 40 to 80\,kpc, we are able to recover between 41 and 67\,per\,cent of the RRLs (in the case of {\it Gaia} and the CRTS, respectively). 
A tentative explanation for this difference, albeit the low number statistics, is the possible contamination in the {\it Gaia} SOS catalogue at these distances, as it has been shown that artefacts and spurious detections might be 
present in crowded areas \citep[e.g., close to the Galactic plane;][]{Holl2018,Clementini2019,Rimoldini2019}, in combination with our survey's photometric completeness limitations and RRL selection strategy. 
Beyond 80\,kpc, no RRLs from the CRTS that fulfill our selection cuts are found, while we recover $\sim$50\,per\,cent of the RRLs from PS-1 and {\it Gaia}.

We find that 99 (98)\,per\,cent of the stars classified as RRab (RRc) in our catalogue have the same classification in the {\it Gaia} catalogue, with a median absolute period difference of 0.0026 (0.0009)\,d. 
Similarly, 99 (100)\,per\,cent of the ab-type (c-type) RRLs identified in our work have the same classification in the CRTS, in which case the median absolute period difference is of 0.0020 (0.0008)\,d. 
Thus, we consider our classification methodology to be robust.

We detect 103, 89, and 87 RRLs that are not listed in the concatenation of the aforementioned catalogues, when crossmatching using a search radius of 1, 3, and 5\,arcsec, respectively.
These stars are located from $\sim$7 to 265\,kpc in heliocentric distance, and the majority of them (58\,per\,cent) lie in the low Galactic latitude fields of the second HOWVAST campaign.  
Interestingly, 80\,per\,cent of the 69 RRLs with $d_{\rm H} \leq $ 80\,kpc are classified as c-type, which might be a consequence of contamination by blended sources (for RRLs near the Galactic plane) and/or the misclassification of variable objects (e.g., eclipsing binaries).  
This does not occur for the stars further than 80\,kpc, where 60\,per\,cent of the new RRLs are classified as ab-type. 
We highlight that the majority of the most distant stars in our catalog are new discoveries. 
We describe this subsample in more detail in Section~\ref{sec:100kpc}.

\vspace{1cm}

\subsection{Classification and Bailey diagram}

In order to classify the RRLs in our catalogue, we adjusted the light curve templates from the SDSS Stripe 82 \citep{Sesar2010} to our phased light curves.
This was performed using the templates in the $g$ and $r$ bands available in the Python package {\sc gatspy} \citep{VanderPlas2015}.
The final classification of RRLs into ab- and c- subtypes was based on the inspection of the best-fitting templates, their amplitudes, and periods.
This resulted in 331 RRab stars (67\,per\,cent), 157 RRc stars (32\,per\,cent), 
and 4 RRLs (1\,per\,cent) that do not fall in either category, with indications of pulsations in the fundamental-mode and first-overtone. 
We classified the latter as RRd stars. 
The relative fractions are similar if we consider the full sample, where 69\,per\,cent and 30\,per\,cent of the stars are classified as RRab and RRc, respectively. 
The distribution of these stars in the period-amplitude space (Bailey diagram) is shown in Figure~\ref{fig:bailey} colour-coded by type.

The position of RRLs in the Bailey diagram can be used to confirm their classification and as a tool to assess the Oosterhoff type \citep[][]{Oosterhoff1939} of stellar systems. 
The Oosterhoff groups are seen as a dichotomy in the mean period and amplitude, and the ratio of RRab and RRc stars in globular clusters.
Cluster RRLs can be split into Oosterhoff-I type (of shorter RRab star periods $\sim$0.55\,d and lower RRc star fractions; OoI), Oosterhoff-II (RRab star periods $\sim$0.65\,d and higher RRc star fractions; OoII), and those that lie in between the two regimes (Oosterhoff-intermediate; Oo-int).
This dichotomy, however, is not present in most MW dwarf galaxies 
and field stars (the latter being predominantly OoI).

Figure~\ref{fig:bailey} shows the OoI, OoII, and Oo-int fiducial lines in the Bailey diagram as defined by \citet{Fabrizio2019} for $V-$band RRL amplitudes. 
To account for the differences between RRL amplitudes in the $g$ and $r-$bands from our work and those in the $V-$band, we scale our amplitudes by a factor of 0.90 and 1.21, respectively \citep{Sesar2012}.
From the orthogonal proximity of the stars to either of the Oosterhoff groups' fiducial lines, we conclude that most of the RRab stars in our sample could be considered OoI or Oo-int, which is the expected trend for field RRLs.
We discuss in more detail the distribution of the most distant RRLs in the Bailey diagram in Section~\ref{sec:100kpc}.

\section{RR Lyrae stars beyond 100\,kpc}
\label{sec:100kpc}

We identify 9 RRLs with mean $g$ and $r$ magnitudes fainter than 20.7 and 20.3, respectively, which corresponds to heliocentric distances larger than 100\,kpc. 
When considering the full sample, the number of RRLs beyond this limit increases to 23. 
The spatial distribution of these stars is depicted with larger symbols in Figure~\ref{fig:skymapRRLs}. 
None of the stars identified in this work are listed in the catalogues used for comparison in Section~\ref{sec:comparison}. 
Among these RRLs, three stars are located beyond 200\,kpc. 
The number of epochs in the light curves of these RRLs spans from 22 to 27 in the $g-$band, and from 15 to 28 in the $r-$band.
The folded light curves of the newly detected RRLs are shown in Figure~\ref{fig:lcs}, and their main properties are summarized in Table~\ref{tab:distant}.

Most stars in this subsample are classified as ab-type (83\,per\,cent), 
which is not surprising given their relative abundance and the fact that they are easier to identify than RRc or RRd. 
The location of these distant RRab stars in the Bailey diagram (Figure~\ref{fig:bailey}), does not show a strong association with the locus of the fiducial line of the OoII group (nor the OoI group), albeit their tendency for periods of pulsation longer than 0.60\,d. 
In fact, the average period of these RRab stars is 0.66\,d, 
similar to the mean period of distant RRLs found by \citet{Medina2018} and the collection of RRLs in MW ultra-faint dwarf galaxies studied by \citet{Viv16} (0.67\,d; see also \citealt{Martinez2021c}).  
Figure~\ref{fig:bailey} shows that the distribution of distant RRLs does not follow the overall trend of RRab stars within 100\,kpc. 
In particular, they are not preferably located near the locus of the OoI group.

High Amplitude Short Period (HASP) variables, that is, those with periods shorter than 0.48\,d, and $V-$band amplitudes larger than 0.75, have been interpreted as coming from progenitors or regions in the Galaxy with populations more metal-rich than [Fe/H] $=-1.5$ \citep{Fiorentino2015}. 
Therefore, RRLs lying in this region of the Bailey diagram can provide insights in the building of the halo and its progenitors. 
In fact, most MW dSphs lack HASP variables, while these stars are not rare in the halo and among globular clusters more metal rich than $-1.5$\,dex and massive dwarf galaxies \citep[][]{Fiorentino2017}. 
In the combined sample, we find 26 RRab stars populating the HASP region, which corresponds to only 
6\,per\,cent of our full RRab star sample. 
The heliocentric distance of these stars (see Section~\ref{sec:distances}) ranges between 9 to 45\,kpc, thus none of the RRLs in the most distant subgroup lie in the HASP region.  
The relatively low fraction of HASP RRLs in our sample might be an indication of the contribution of dwarf galaxies to the dual origin of the outer halo (and its dependence on Galactocentric distance), as further discussed in Section~\ref{sec:density}. 
We note, however, that additional evidence is required to support this assertion.

\begin{table*}
\small
\caption{
Main properties of the potential groups of RRLs beyond 100\,kpc discussed in Section~\ref{sec:groups}. 
}
\label{tab:groups}
\begin{center}
\begin{tabular}{|c|c|c|c|c|c|c|c|c|c|}

\hline

ID &           R.A. &          DEC. &       Period & Amplitude & Band & Type & $d_{\rm H}$ & $R_{\rm sph}$ & $R_{\rm el}$ \\
  & (deg) & (deg) & (days ) & & & & (kpc) & (kpc) & (kpc)     \\
\hline
\multicolumn{10}{|c|}{Group 1}\\
\hline
   HV153403-321831 &  $233.51328$ &  $-32.30856$ &  $0.8080$ &    $0.66$ & $r$ &   ab &   $135\pm5$ &    $128\pm4$ &    $136\pm5$ \\
   HV152905-315335 &  $232.27196$ &  $-31.89316$ &  $0.8129$ &    $0.47$ & $r$ &   ab &   $144\pm5$ &    $137\pm5$ &    $146\pm5$ \\
 \hline

\multicolumn{10}{|c|}{Group 2}\\
\hline
 HV154834-320810 &  $237.14125$ &   $-32.13607$ &  $0.5967$ &    $0.38$ &  $r$ &   ab &   $110\pm4$ &    $103\pm3$ &    $108\pm4$ \\
 HV155407-361645 &  $238.52769$ &   $-36.27926$ &  $0.6618$ &    $0.65$ & $r$ &  ab &   $104\pm4$ &    $96\pm3$ &    $99\pm4$ \\
\hline
\end{tabular} 
 \vspace{0.001cm}
\end{center}
\end{table*}

To examine the overall consistency of the number of distant RRLs found in our work with the results of previous studies of similar photometric depth, we can perform a direct comparison assuming high completeness out to similar distances. 
\citet{Stringer21} detected 6,971 ab-type RRL candidates in the $\sim$5,000\,deg$^2$ of the DES' footprint, among which 4,569 do not belong to the known substructures and galaxies considered by the authors. 
Of this subsample, 18\,per\,cent 
are located beyond 100\,kpc from the Sun, which implies a rough density of six 
distant halo RRLs every 40\,deg$^2$ (or 0.16\,per deg$^2$), without accounting for their estimated completeness (expected to be $>70$\,per\,cent at $\sim150$\,kpc). 
This number is a factor of three larger than the number of distant RRLs 
detected in our study (two RRab stars beyond 100\,kpc every 40\,deg$^2$, or 0.056 per deg$^2$). 
Nonetheless, our density is more consistent with the findings of \citet{Stringer21} if we only consider their candidates with more than 25 observations in total (considering $g$, $r$, $i$, $z$, and $Y$) and with an RRab score $>0.90$ as assigned by their classifier. 
In this case, the DES RRab star density decreases to 0.083 distant RRLs per square degree (or three RRLs every 40\,deg$^2$), in rough agreement with our results.

\subsection{Potential groups of distant RR Lyrae stars}
\label{sec:groups}

In Figure~\ref{fig:skymapRRLs}, we highlight the spatial distribution of the RRL candidates detected beyond 100\,kpc. 
From the figure we identify two groups of stars with similar on-sky positions and heliocentric distances. 
Associations and groups at large distances (especially at $d_{\rm H}>$ 100\,kpc) are unlikely to happen by chance from halo stars. 
Using mock stellar haloes and focusing on RRLs beyond 100\,kpc, \citet{Sanderson2017} 
showed that the median of the minimum angular distance to the nearest star for bound (unbound) RRLs beyond 100\,kpc is $\sim$0.01\,deg (3.0\,deg), with the majority of minimum separations between 0--0.03\,deg ($\sim$0.3--10.0\,deg).
Moreover, \citet{Sanderson2017} found that the closest pairs of RRLs tend to originate from the same building block (regardless of their bound/unbound status).

We analyzed these groups looking for indications of their potential association with known substructures. 
The main properties of the RRLs in these groups are presented in Table~\ref{tab:groups}. 
The first group consists of the stars HV153403-321831 and HV152905-315335, two RRab stars located at $135\pm5$ and $144\pm5$\,kpc, respectively (with right ascensions of $\sim$233\,deg). 
These stars are separated by 1.1\,deg, which corresponds to $\sim$3\,kpc at $d_{\rm H}\sim$ 140\,kpc. 
The second group, at right ascensions of $\sim$237.5\,deg, comprises the stars HV154834-320810 and HV155407-361645, with distances of $110\pm4$ and $104\pm4$, respectively.
Both of these stars are classified as RRab stars. 
This group shows an angular extension of $\sim$4.3\,deg (or 8\,kpc at $d_{\rm H}\sim$ 107\,kpc). 
We note that the stars in both groups are too separated to be considered part of an intact (or not heavily disrupted) bound satellite, but they lie well within the predicted range of separations for unbound (but associated) debris from \citealt{Sanderson2017}.  

As these RRLs might be associated with the ongoing tidal disruption of MW satellites, we inspected the Python library {\it galstreams} \citep[][]{Mateu2022}, which collects celestial, distance, proper motion, and radial velocity information for $\sim$97 distinct stellar streams.   
Nevertheless, we find no streams within the {\it galstreams} database with distances similar to those of our groups (most of the streams close to the position of our groups have $d_{\rm H}<$ 40\,kpc). 
Comparing the positions and distances of these groups to the model of the Sagittarius stream by \citet{Dierickx2017} shows that only the group with higher right ascension is in proximity to the stream, but at larger distances (most Sagittarius stream stars at this right ascension are located at $d_{\rm H}\sim$ 50\,kpc ).
Moreover, all of the stars in these groups have latitude-like coordinates in the Sagittarius stream system ($B_{\rm Sgr}$; \citealt{Majewski2003}) larger than 16.6\,deg, making their association with the stream unlikely. 
Therefore, we find no clear indications of associations between our groups of clumped RRLs and known satellites or streams.  
Nevertheless, we suggest that the association of the stars in these groups is likely.

\section{Space density distribution}
\label{sec:density}

\subsection{Radial density model}
\label{sec:models}

\begin{figure*}

\begin{center}
\includegraphics[angle=0,scale=.22]{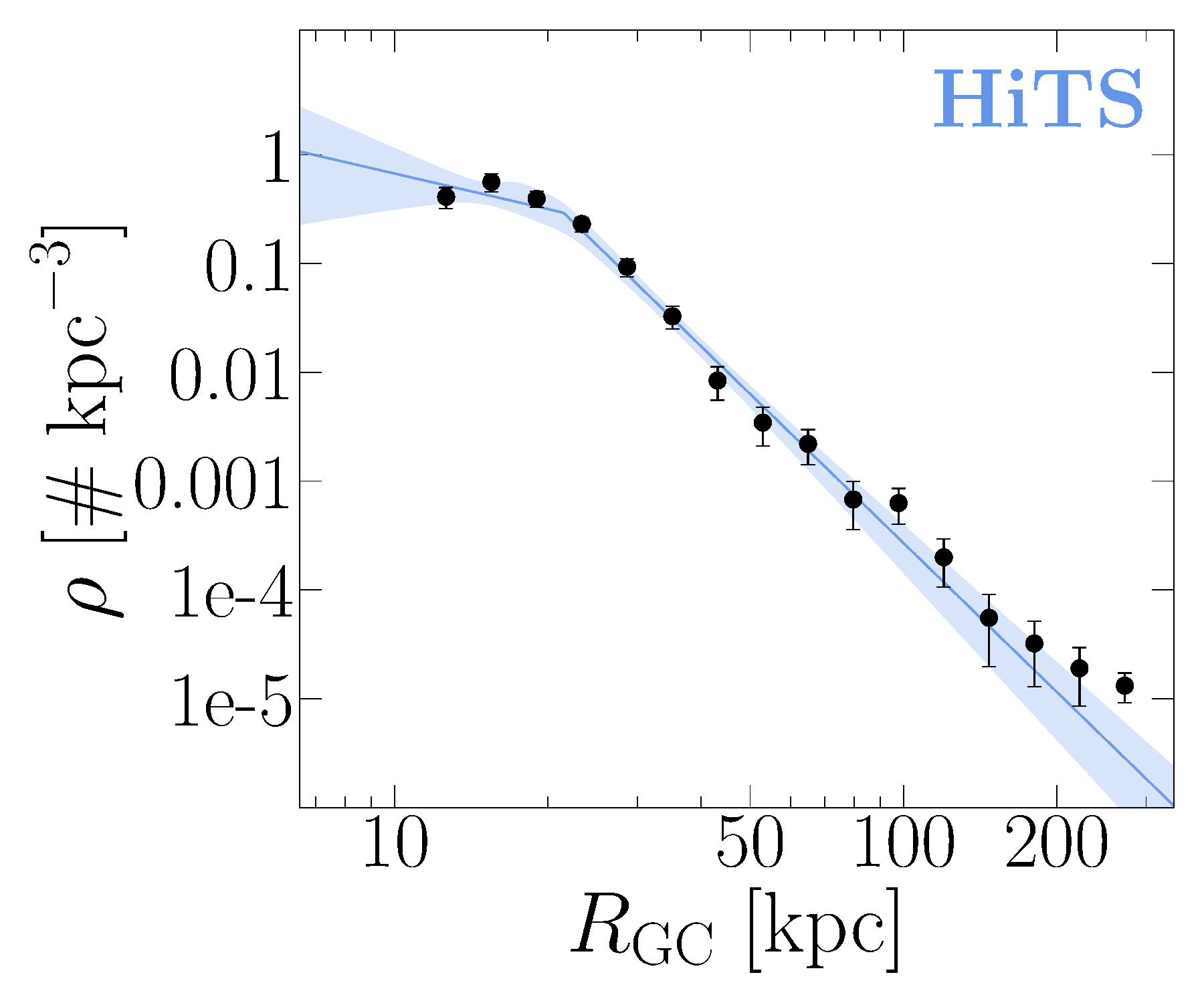}
\includegraphics[angle=0,scale=.22]{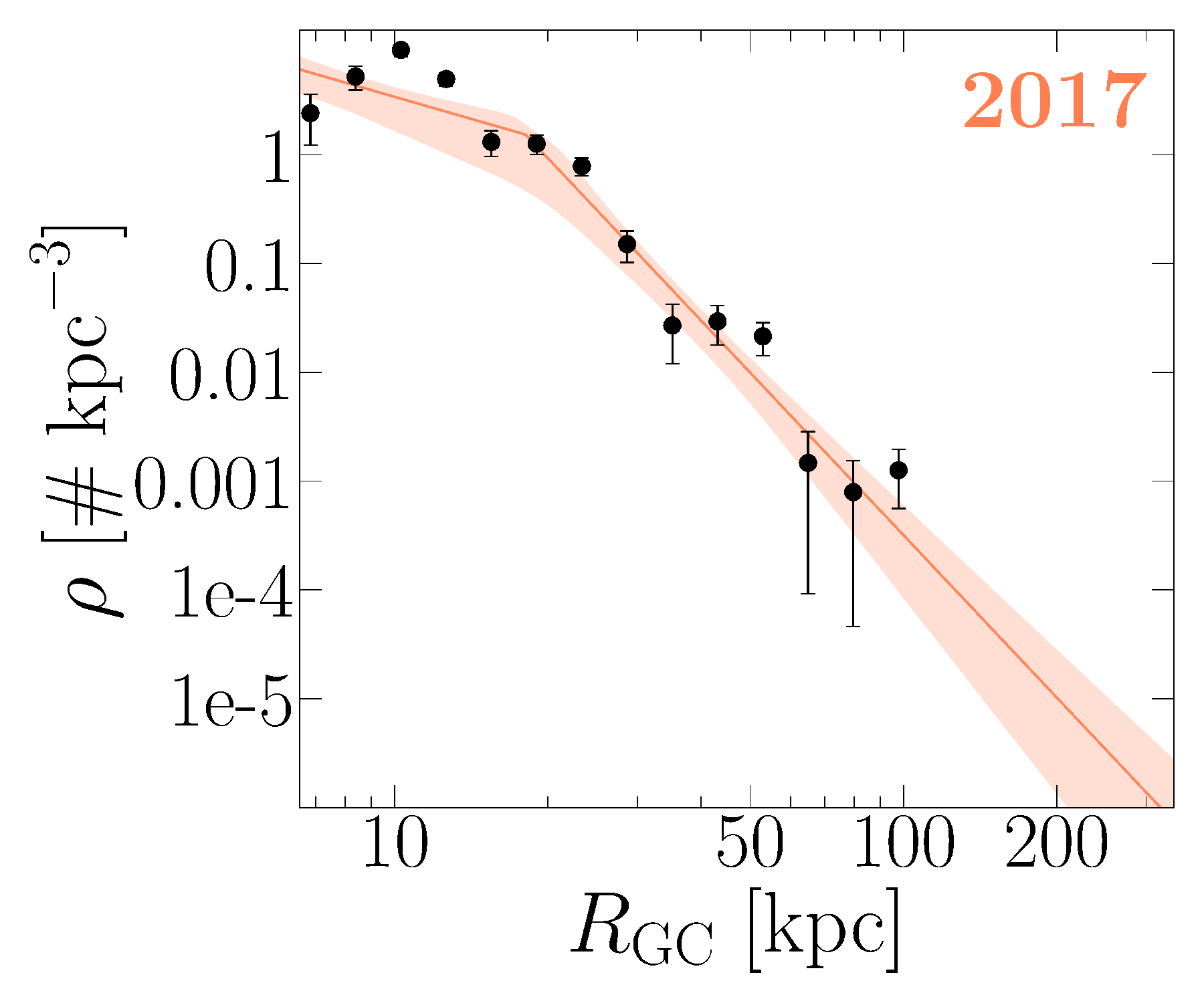}
\includegraphics[angle=0,scale=.22]{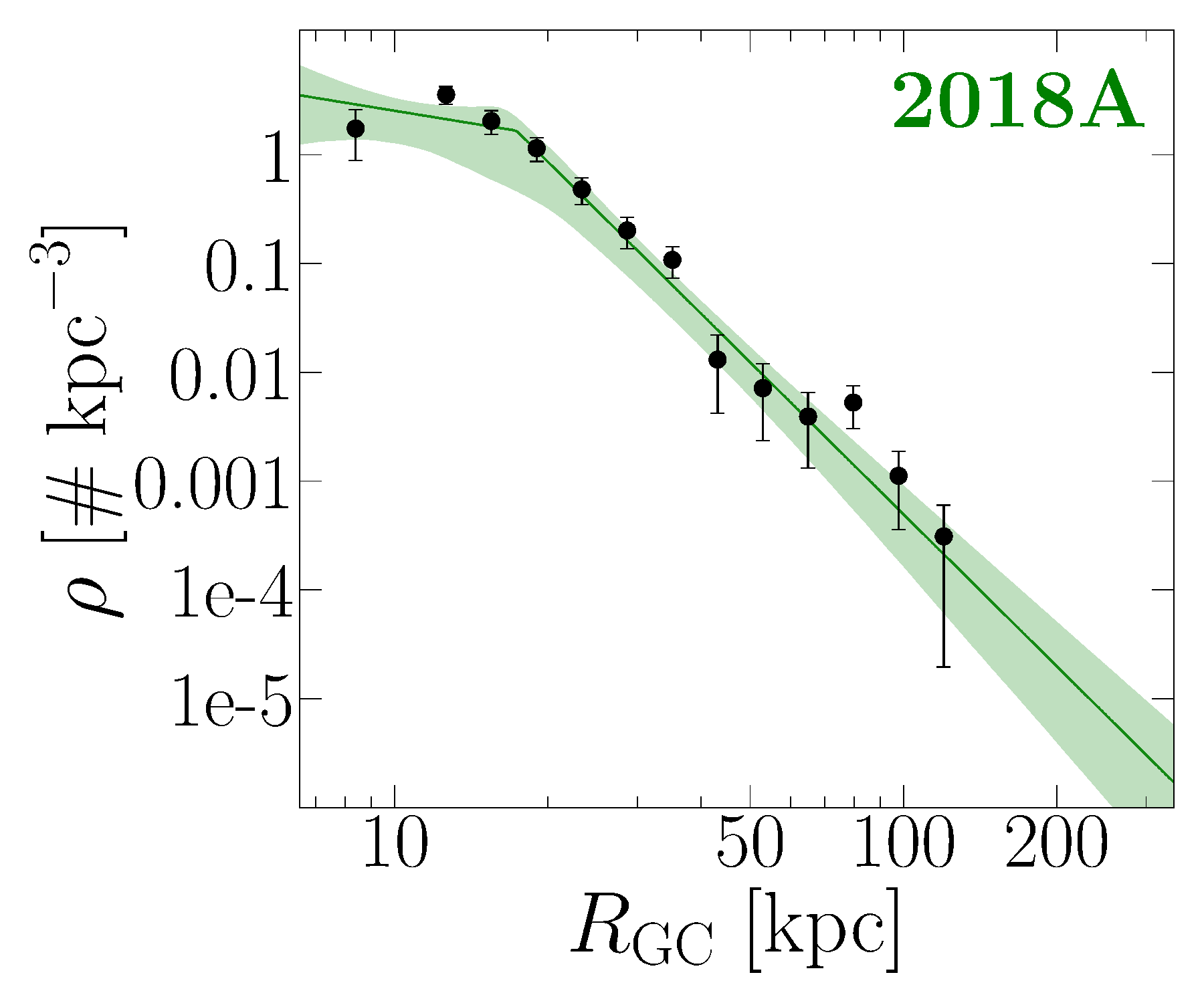}
\includegraphics[angle=0,scale=.22]{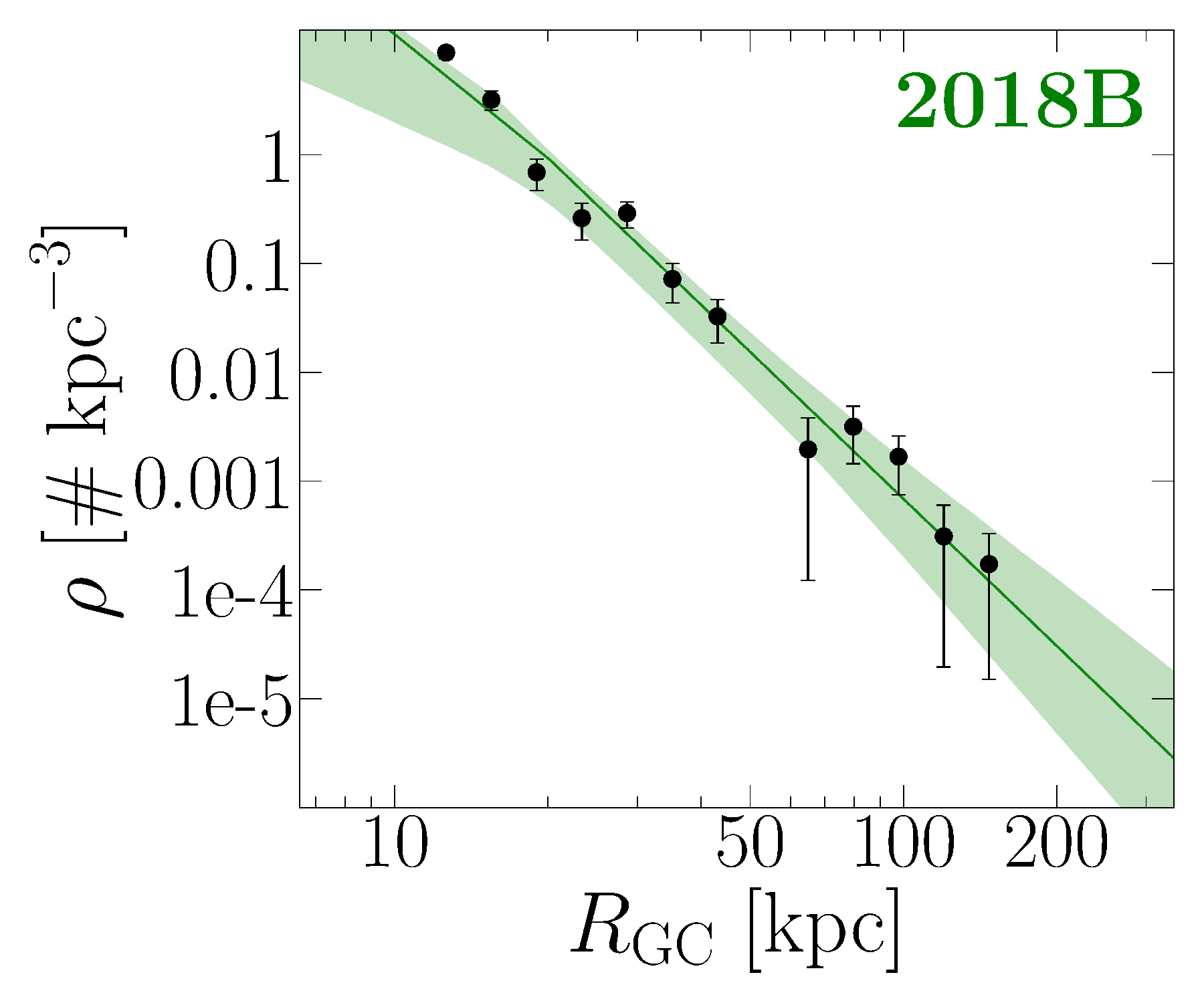}
\includegraphics[angle=0,scale=.25]{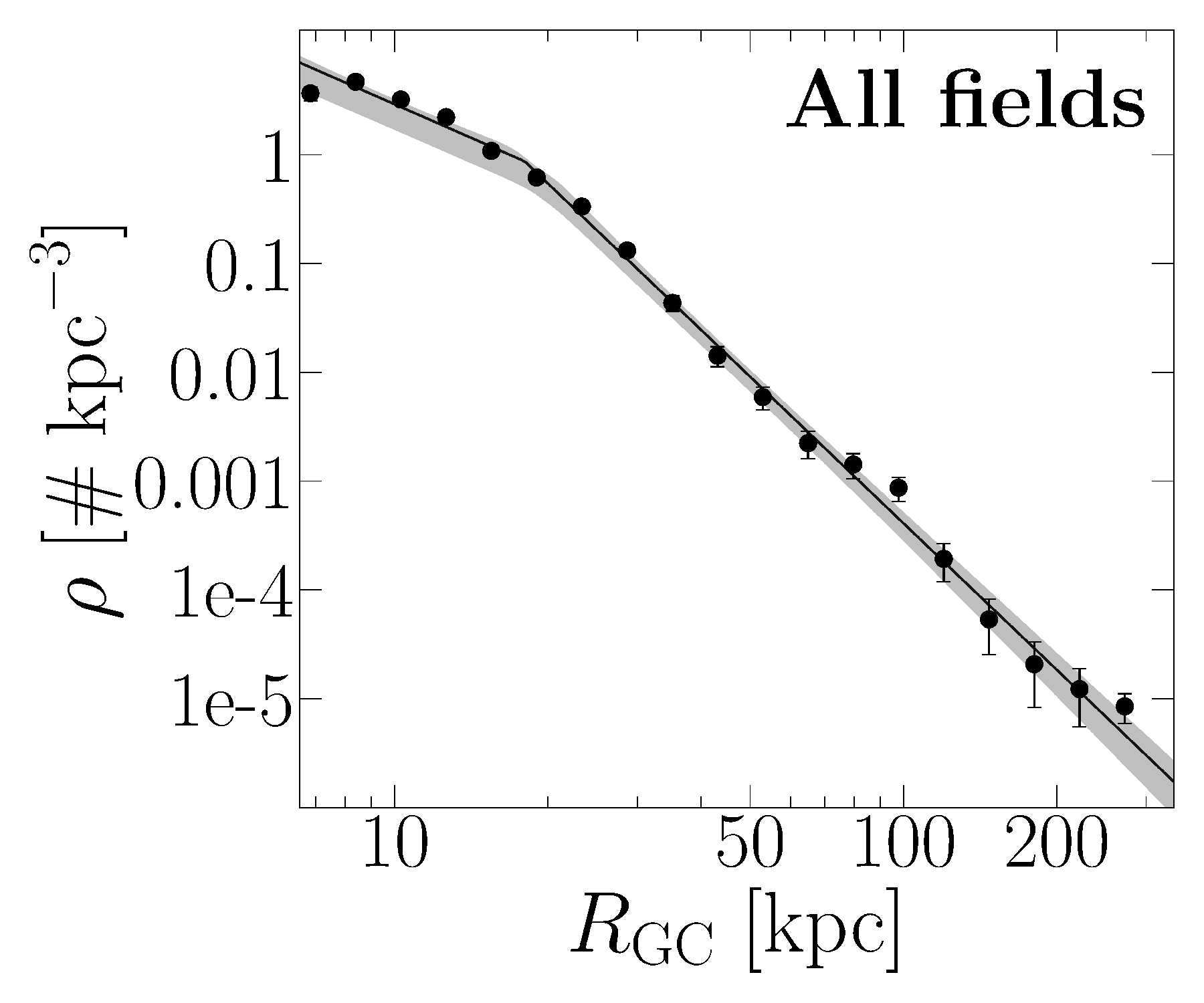}
\caption{
Binned RRL number density profiles of the regions studied in this work, for a spherical halo model.  
The region corresponding to the HiTS fields is shown with blue curves, whereas that of HOWVAST 2017 is shown in orange.
Green curves depict the high and low Galactic latitude areas surveyed by HOWVAST 2018 ({\it left} and {\it right}, respectively).
The profile resulting from considering the entire studied region is shown in black.
The best solution determined via MCMC is shown as a solid line in each panel, and the shaded regions depict the $3\sigma$ confidence levels.
The uncertainty shown for each density bin represents Poisson noise. 
}
\label{fig:dens}
\end{center}
\end{figure*}

In this section we address in more detail the radial density distribution of our full sample. 

Many observational studies and simulations have suggested that the properties of the radial distribution of stars in the halo is connected with their origin \citep[e.g.,][]{Zinn93,Vivas2006,Pillepich2014}. 
From the slope(s) of the radial distribution, for instance, one can infer the existence of an inner halo thought to contain both accreted and formed in-situ stars, and an outer halo, expected to have been formed largely from the accretion of satellites \citep[e.g.,][]{Wat09,BJ05,Zolo09,Naidu2020}

In order to characterize the spatial distribution of our RRLs in the halo, we follow the methodology of \citet{Medina2018}.
We adopt two models to represent the data -- one assuming a spherical halo, and one for an ellipsoidal halo.
To account for the oblateness of the latter, we assume $q=c/a=0.7$ \citep{Sesar2011}, where $a$ and $c$ are the axes in the disc plane and along the vertical direction, respectively. 
Thus, prior to binning our sample in distance, we transform our computed distances from heliocentric to spheroidal and ellipsoidal Galactocentric distances ($R_{\rm sph}$ and $R_{\rm el}$) using: 

\begin{equation}
\begin{array}{cc}
R_{\rm sph}^2=(R_\odot - d_{\rm H} \cos{b} \cos{l})^2 + d_{\rm H}^2 \cos^2{b} \sin^2{l} + d_{\rm H}^2 \sin^2{b}\\
R_{\rm el}^2=(R_\odot - d_{\rm H} \cos{b} \cos{l})^2 + d_{\rm H}^2 \cos^2{b} \sin^2{l} + (d_{\rm H}/0.7)^2 \sin^2{b},
\end{array}
\label{eq:Rgc}
\end{equation}

\noindent where $R_\odot$ stands for the distance from the Sun to the Galactic centre (assumed to be 8\,kpc; \citealt{Gravity2021}), and $b$ and $l$ are the Galactic latitude and longitude, respectively.
The adoption of two fixed flattening parameters is motivated by our limitations in constraining the shape of the halo from our survey’s small (and non-contiguous) footprint, which would lead to over-interpretations of our results.

We adopt a power-law model to describe the radial density $\rho(R_{\rm GC})$ of our halo RRLs, where $R_{\rm GC}$ represents either $R_{\rm sph}$ or $R_{\rm el}$.
Thus, $\rho(R_{\rm GC}) = \rho_\odot(R_{\rm GC}/R_\odot)^n$, where  $\rho_\odot$ is the local RRL number density and $n$ is the slope of the profile for a simple power law.
Given that vast observational evidence suggest the existence of a break in the halo radial density profile between 20 and 35\,kpc \citep[e.g.,][]{Saha85,Wat09,Deas11,Sesar2011,Belokurov2018a,Stringer21}, we describe the explored regions with simple and broken power-law models. 
For the latter, a break at $R_{\rm break}$ is used, so that, in logarithmic form:

\begin{equation}
\begin{array}{ccc}
\log{(\rho (R_{\rm GC}))}=A_1+n_1\ \log{(R_{\rm GC}/R_\odot)}\\
\log{(\rho (R_{\rm GC}))}=A_2+n_2\ \log{(R_{\rm GC}/R_\odot)}\\
A_1+n_1\ \log{(R_{\rm break}/R_\odot)}=A_2+n_2\ \log{(R_{\rm break}/R_\odot)}
\end{array}
\label{eq:logarithm_2}
\end{equation}

\noindent where $A = \log{(\rho_\odot)}$, and the subindices denote each side of the density profile (i.e., $A_1$ and $A_2$ correspond to the inner and outer density, respectively). 

To explore the parameter space and their distribution, we employ {\it emcee} \citep{Foreman2013}, a Python implementation of the invariant Markov chain Monte Carlo method (MCMC).
For this, we leave $A_2$, $n_1$, $n_2$, and  $R_{\rm break}$ as free parameters, and adopt the priors used by \citet{Medina2018}.
We find that running {\it emcee} with 200 walkers and a chain of 500 steps is sufficient to reach convergence.
The selected values are provided in Table~\ref{tab:densparams} and correspond to the median of the marginalized posterior parameter distributions, and their errors represent their 16th and 84th percentiles. 
The posterior probability distribution for the obtained parameters of the broken-power-law model of the oblate halo 
are depicted in Figure~\ref{fig:corner1}. 
We examine our findings and their implications in the following section.

\begin{table*}
\small
\caption{
Simple and broken-power-law parameters from the sampled posterior probability distributions described in Section~\ref{sec:density}. We report the results from using the RRLs in all of our regions combined, and from the individual areas of the survey, as well as the adopted shape of the halo (spheroidal and oblate, denoted as $R_{\rm sph}$ and $R_{\rm el}$, respectively). 
For most of the studied regions the best-fitting model corresponds to the broken-power-law profiles.  
}
\label{tab:densparams}
\begin{center}

\begin{tabular}{|c|c|HHHc|c|c|c|c|c|c|c|c|}
\hline
                Region & $R_{\rm GC}$ type &    Mask & Corrected & RRL Types &                      $A$ &                        $n$ & Simple $\chi^2_{\nu} $ &                    $A_1$ &                   $A_2$ &                    $n_1$ &                    $n_2$ &  $R_{\rm break}$ & Broken $\chi^2_{\nu} $ \\

                & &    &  &                  &     &                         &  &               &                  &                     &                  &  (kpc) & \\
\hline
\multicolumn{14}{|c|}{All RRLs}\\
\hline
   All &  Rgc & Mask &      Corr &  $AllRRL$ & $0.69^{+0.02}_{-0.02}$ & $-3.42^{+0.03}_{-0.03}$ &                $0.058$ &  $0.67^{+0.02}_{-0.03}$ & $1.52^{+0.13}_{-0.10}$ & $-2.05^{+0.13}_{-0.15}$ & $-4.47^{+0.11}_{-0.18}$ &  $18.0^{+2.1}_{-1.1}$ &       $0.006$ \\
   All &  Rel & Mask &      Corr &  $AllRRL$ & $0.72^{+0.02}_{-0.02}$ & $-3.42^{+0.03}_{-0.03}$ &                $0.038$ &  $0.67^{+0.02}_{-0.05}$ & $1.69^{+0.20}_{-0.19}$ & $-2.47^{+0.28}_{-0.12}$ & $-4.57^{+0.17}_{-0.25}$ &  $24.3^{+2.6}_{-3.2}$ &       $0.011$ \\
  2017 &  Rgc & Mask &      Corr &  $AllRRL$ & $0.58^{+0.05}_{-0.06}$ & $-3.24^{+0.08}_{-0.09}$ &                $7.806$ &  $0.66^{+0.06}_{-0.09}$ & $1.94^{+0.32}_{-0.37}$ & $-1.35^{+0.33}_{-0.51}$ & $-4.96^{+0.51}_{-0.52}$ &  $18.1^{+2.1}_{-1.5}$ &       $2.908$ \\
  2017 &  Rel & Mask &      Corr &  $AllRRL$ & $0.47^{+0.06}_{-0.07}$ & $-2.82^{+0.09}_{-0.09}$ &                $1.080$ &  $0.37^{+0.09}_{-0.10}$ & $1.71^{+0.48}_{-0.31}$ & $-1.00^{+0.58}_{-0.64}$ & $-4.41^{+0.39}_{-0.60}$ &  $20.7^{+3.1}_{-3.8}$ &       $0.456$ \\
 2018A &  Rgc & Mask &      Corr &  $AllRRL$ & $0.66^{+0.08}_{-0.09}$ & $-3.24^{+0.12}_{-0.12}$ &                $3.700$ &  $0.45^{+0.13}_{-0.16}$ & $1.78^{+0.23}_{-0.26}$ & $-0.76^{+0.87}_{-1.04}$ & $-4.64^{+0.31}_{-0.41}$ &  $17.4^{+2.7}_{-1.4}$ &       $1.116$ \\
 2018A &  Rel & Mask &      Corr &  $AllRRL$ & $0.49^{+0.09}_{-0.11}$ & $-3.12^{+0.13}_{-0.14}$ &                $0.990$ &  $0.01^{+0.29}_{-0.63}$ & $1.78^{+0.38}_{-0.39}$ & $-0.55^{+2.66}_{-1.36}$ & $-4.57^{+0.39}_{-0.58}$ &  $21.3^{+4.2}_{-3.4}$ &       $0.635$ \\
 2018B &  Rgc & Mask &      Corr &  $AllRRL$ & $1.50^{+0.02}_{-0.03}$ & $-4.10^{+0.07}_{-0.08}$ &               $33.446$ &  $1.50^{+0.03}_{-0.20}$ & $1.76^{+0.24}_{-0.42}$ & $-3.78^{+0.57}_{-0.31}$ & $-4.49^{+0.47}_{-0.47}$ & $20.1^{+14.6}_{-4.7}$ &      $23.599$ \\
 2018B &  Rel & Mask &      Corr &  $AllRRL$ & $1.58^{+0.02}_{-0.02}$ & $-4.19^{+0.06}_{-0.07}$ &               $14.589$ &  $1.58^{+0.03}_{-0.35}$ & $1.92^{+0.23}_{-0.61}$ & $-3.64^{+0.61}_{-0.52}$ & $-4.70^{+0.71}_{-0.51}$ & $17.9^{+15.0}_{-2.2}$ &       $7.515$ \\
  HiTS &  Rgc & Mask &      Corr &  $AllRRL$ & $0.52^{+0.06}_{-0.06}$ & $-3.48^{+0.09}_{-0.10}$ &                $0.037$ & $-0.07^{+0.24}_{-0.25}$ & $1.42^{+0.21}_{-0.17}$ & $-1.08^{+0.93}_{-0.83}$ & $-4.55^{+0.22}_{-0.31}$ &  $21.5^{+2.0}_{-2.1}$ &       $0.007$ \\
  HiTS &  Rel & Mask &      Corr &  $AllRRL$ & $0.04^{+0.04}_{-0.02}$ & $-3.11^{+0.07}_{-0.09}$ &                $0.105$ & $-0.50^{+0.31}_{-0.69}$ & $1.30^{+0.31}_{-0.26}$ & $-0.99^{+2.31}_{-1.01}$ & $-4.39^{+0.29}_{-0.38}$ &  $26.2^{+4.6}_{-4.6}$ &       $0.045$ \\

\hline
\multicolumn{14}{|c|}{ab-type only}\\
\hline
   All &  Rgc & Mask &      Corr &  $abOnly$ & $0.47^{+0.02}_{-0.02}$ & $-3.37^{+0.04}_{-0.04}$ &                $0.064$ &  $0.45^{+0.03}_{-0.04}$ & $1.45^{+0.16}_{-0.12}$ & $-1.88^{+0.23}_{-0.16}$ & $-4.59^{+0.18}_{-0.23}$ & $18.7^{+1.9}_{-1.5}$ &      $0.006$ \\
   All &  Rel & Mask &      Corr &  $abOnly$ & $0.47^{+0.02}_{-0.02}$ & $-3.31^{+0.04}_{-0.04}$ &                $0.037$ &  $0.43^{+0.03}_{-0.04}$ & $1.53^{+0.24}_{-0.18}$ & $-2.19^{+0.14}_{-0.14}$ & $-4.56^{+0.18}_{-0.31}$ & $23.3^{+3.1}_{-2.0}$ &      $0.005$ \\
  2017 &  Rgc & Mask &      Corr &  $abOnly$ & $0.55^{+0.05}_{-0.06}$ & $-3.05^{+0.10}_{-0.10}$ &               $15.110$ &  $0.56^{+0.06}_{-0.08}$ & $1.74^{+0.38}_{-0.30}$ & $-1.52^{+0.38}_{-0.42}$ & $-4.80^{+0.42}_{-0.67}$ & $18.1^{+2.9}_{-1.7}$ &     $10.860$ \\
  2017 &  Rel & Mask &      Corr &  $abOnly$ & $0.41^{+0.07}_{-0.08}$ & $-3.06^{+0.11}_{-0.12}$ &                $0.852$ &  $0.30^{+0.09}_{-0.11}$ & $1.94^{+0.48}_{-0.47}$ & $-1.34^{+0.43}_{-0.51}$ & $-4.86^{+0.49}_{-0.71}$ & $23.1^{+2.8}_{-3.2}$ &      $0.368$ \\
 2018A &  Rgc & Mask &      Corr &  $abOnly$ & $0.60^{+0.09}_{-0.11}$ & $-3.35^{+0.15}_{-0.15}$ &                $2.254$ &  $0.42^{+0.13}_{-0.17}$ & $1.70^{+0.32}_{-0.29}$ & $-1.18^{+0.90}_{-0.79}$ & $-4.80^{+0.40}_{-0.56}$ & $17.6^{+3.1}_{-1.6}$ &      $0.683$ \\
 2018A &  Rel & Mask &      Corr &  $abOnly$ & $0.44^{+0.11}_{-0.13}$ & $-3.23^{+0.17}_{-0.17}$ &                $0.610$ &  $0.12^{+0.21}_{-0.39}$ & $1.76^{+0.67}_{-0.46}$ & $-1.21^{+1.57}_{-0.87}$ & $-4.86^{+0.59}_{-1.02}$ & $22.0^{+4.5}_{-3.6}$ &      $0.288$ \\
 2018B &  Rgc & Mask &      Corr &  $abOnly$ & $1.24^{+0.03}_{-0.03}$ & $-3.99^{+0.10}_{-0.11}$ &                $0.495$ &  $1.24^{+0.04}_{-0.17}$ & $1.74^{+0.36}_{-0.35}$ & $-3.44^{+0.57}_{-0.40}$ & $-4.88^{+0.54}_{-0.85}$ & $17.8^{+7.8}_{-2.0}$ &      $0.202$ \\
 2018B &  Rel & Mask &      Corr &  $abOnly$ & $1.31^{+0.03}_{-0.03}$ & $-4.02^{+0.08}_{-0.08}$ &                $1.442$ &  $1.32^{+0.04}_{-0.11}$ & $1.89^{+0.23}_{-0.33}$ & $-3.28^{+0.35}_{-0.31}$ & $-5.05^{+0.59}_{-0.57}$ & $17.1^{+3.8}_{-1.4}$ &      $0.480$ \\
  HiTS &  Rgc & Mask &      Corr &  $abOnly$ & $0.44^{+0.07}_{-0.07}$ & $-3.53^{+0.10}_{-0.10}$ &                $0.031$ & $-0.13^{+0.24}_{-0.28}$ & $1.29^{+0.20}_{-0.16}$ & $-1.21^{+1.04}_{-0.80}$ & $-4.50^{+0.21}_{-0.30}$ & $21.6^{+3.0}_{-2.6}$ &      $0.004$ \\
  HiTS &  Rel & Mask &      Corr &  $abOnly$ & $0.04^{+0.04}_{-0.02}$ & $-3.17^{+0.07}_{-0.09}$ &                $0.074$ & $-0.58^{+0.37}_{-0.37}$ & $1.27^{+0.29}_{-0.24}$ & $-0.90^{+1.28}_{-1.13}$ & $-4.40^{+0.25}_{-0.38}$ & $26.3^{+5.1}_{-4.2}$ &      $0.028$ \\
\hline

\end{tabular} 
\end{center}
\end{table*}

\subsection{Number density profiles}

We report the results of the aforementioned methodology for our catalogue excluding only the subsample of stars belonging to the Sextans dSph.
Based on the estimations of our detection efficiency and completeness, we constrain our analysis to RRLs with distances smaller than $145$\,kpc, the limit at which the idealized RRL recovery rate is $\sim$80\,per\,cent and beyond which this rate decreases drastically.
Additionally, we correct our densities by the RRL detection completeness values obtained in Section~\ref{sec:completeness}. 
While our RRL recovery rates from comparisons to other surveys are somewhat lower than the completeness expected based on our idealized estimates, we use the completeness values based on our dataset. 
We choose this because our data are deeper than any of the literature surveys, and thus expected to be more complete. 
Furthermore, we wish to avoid the complicating effects of unknown rates of spurious RRL detections and incompleteness in existing surveys. 
In order to inspect the (in-)homogeneity of the RRLs radial distribution, we follow two approaches: one computing the density profile of our entire sample of RRLs, and one measuring the profile in the four distinct regions covered by our survey. 
By doing this we could, in principle, directly inspect anisotropies in the halo distribution of stars at relatively small scales,  albeit with the additional challenge of low-number statistics. 
Lastly, for each region, we also included an analysis of the distribution of ab-type RRLs only, as they are typically less affected by contamination and are more uniquely identifiable than c-type and d-type RRLs.

\begin{figure}
\begin{center}
\includegraphics[width=0.49\textwidth]{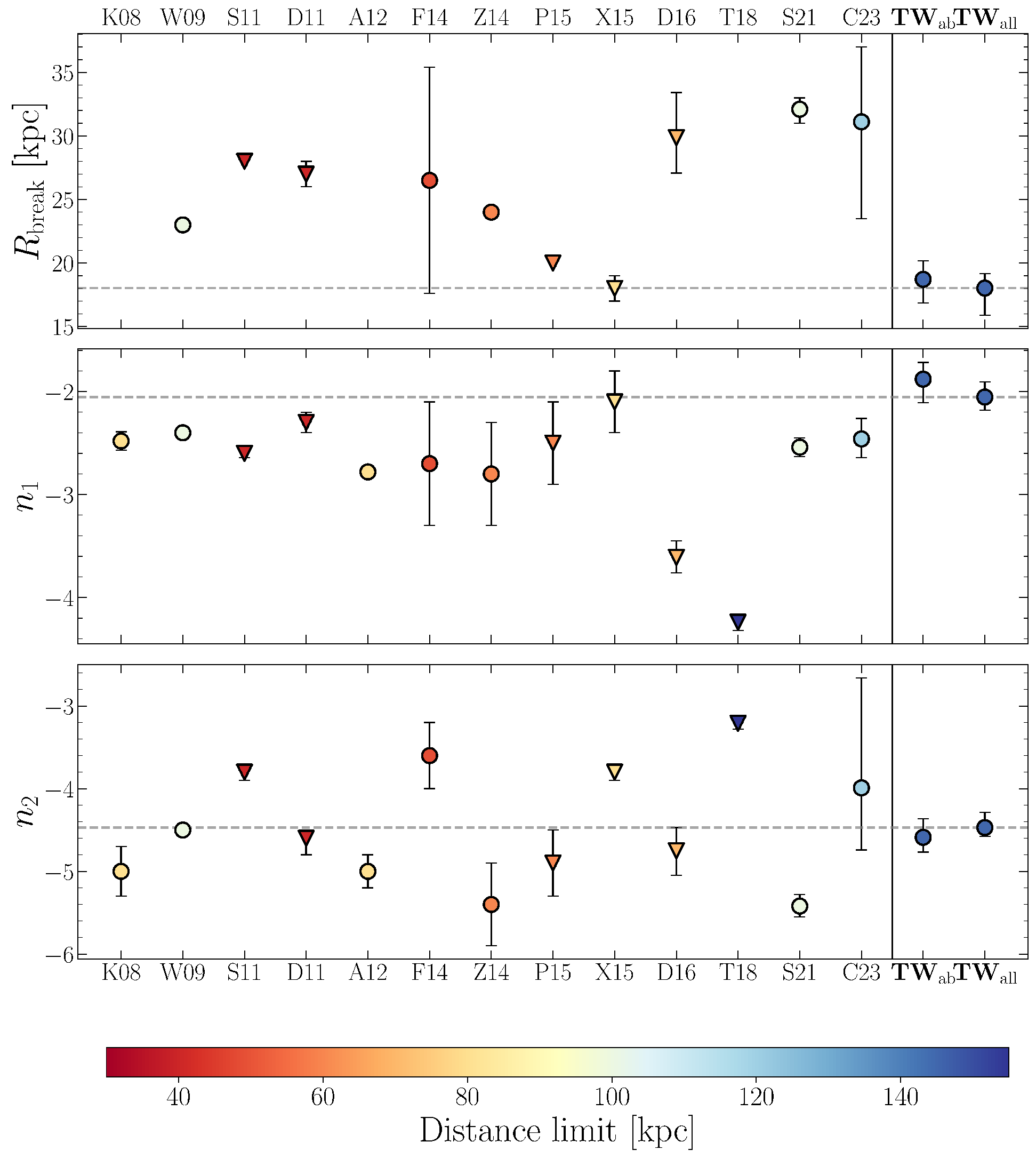}
\caption{
Comparison between our best-fitting radial density profile parameters for a BPL spherical halo model, and those from other BPL halo studies in the literature. 
Parameters obtained from the spatial distribution of RRLs only are shown with circles, whereas triangles represent those from studies based on different halo tracers (e.g., BHB stars or K-giants). 
The markers are colour-coded to illustrate the distance limits of each work, and the error bars depict the uncertainties in the parameters estimation, when available.   
We denote our values as TW$_{\rm ab}$ for the parameters from our RRab-only MCMC model, and TW$_{\rm all}$ for the model based on our full RRL sample. 
A horizontal dashed line is used to highlight the results of the latter. 
The references for the literature works are defined as follows: K08 \citep{Keller2008}, W09 \citep{Wat09}, S11 \citep{Sesar2011}, D11 \citep{Deas11}, A12 \citep{Akhter2012}, F14 \citep{Faccioli2014}, Z14 \citep{Zinn14}, P15 \citep{PilaDiez2015}, X15 \citep{Xue2015}, D16 \citep{Das2016}, T18 \citep{Thomas2018}, S21 \citep{Stringer21}, and C23 \citep{Chen2023}.   
We note that the $R_{\rm break}$ values found by K08 (45\,kpc), A12 (45\,kpc), and T18 (41.4\,kpc) are not shown in the upper panel, for illustrative purposes.  }
\label{fig:comparison}
\end{center}
\end{figure}

Our results for a spherical 
halo model and no RRL class distinction 
are shown in Figure~\ref{fig:dens}, and a summary of the overall results is provided in Table~\ref{tab:densparams}. A first look into Figure~\ref{fig:dens} and Table~\ref{tab:densparams} reveals that, regardless of the region considered, a break is visible between $15$ and $25$\,kpc. 
This feature is, however, less clear in the fields closer to the Galactic plane (2017 and 2018B, see Figure~\ref{fig:skymap}), where more RRLs are detected overall at distances $< 20$\,kpc. 
This is also the case if we restrict our sample to ab-type RRLs only. 
Because ab-type RRLs are less prone to contamination and the vast majority of the stars in our sample have Galactic coordinates $|z|>2$\,kpc, the observed difference is likely due to the overall higher density of stars towards the Galactic plane (as a reference, \citealt{Mateu2018b} estimated the thick disc scale height to be $h_z\sim0.65$\,kpc using RRLs). 
Based on the resulting reduced $\chi^2$ ($\chi^2_\nu$) values of the models, we conclude that broken-power-law models yield better fits than simple-power-laws. 
The break in the profile is also clearly observed in the plot of Figure~\ref{fig:dens}, depicting the radial distribution of RRLs when using all the regions combined. 
In the latter, we see that the underdensities and overdensities at given distance bins visible in the individual region profiles get averaged out, as a consequence of better number statistics. 
In that case, the best-fit case corresponds to the spherical halo and we find $R_{\rm break}$ at $18.1^{+2.1}_{-1.1}$\,kpc, where the profile displays an inner slope of $n_1=-1.88^{+0.23}_{-0.16}$ and a steeper outer halo slope of $n_2=-4.47^{+0.11}_{-0.18}$.

We find that the break radii from the different regions are consistent within their uncertainties (Table~\ref{tab:densparams}), whereas $n_2$ displays a larger dispersion overall.
For a spherical halo, these values vary between $17.4^{+2.7}_{-1.4}$ and $21.5^{+2.0}_{-2.1}$\,kpc for the $R_{\rm break}$, and $-4.96^{+0.51}_{-0.52}$  and $-4.49^{+0.47}_{-0.47}$  for the outer halo slopes.
The largest differences are seen when contrasting the inner slopes $n_1$, which tend to be steeper for the fields near the Galactic plane. 

In Table~\ref{tab:powerLaw}, we display the slopes of the radial distribution of different tracers reported in the literature, and in Figure~\ref{fig:comparison} we contrast our best-fit parameters with broken-power-law fits from these works. 
These works generally model the radial distribution adopting (or deriving different halo flattenings, and cover large areas and different regions of the halo.   
In this regard, we note that our survey maps a smaller area than previous studies, but we are able to better trace the halo beyond 40\,kpc. 
We find that our measured break radii are consistently smaller than those found by \citet{Stringer21}, \citet{Das2016}, and \citet{Chen2023} (who find $R_{\rm break}$ closer to $\sim$30\,kpc), regardless of the adopted halo flattening, and are most similar to those from \citet{PilaDiez2015} and \citet{Xue2015}
(which are compatible with the transition radius between the inner and outer halo in models of hierarchical Galaxy formation; see e.g. \citealt{Tissera2014}). 
This can be interpreted as a direct consequence of the halo flattening adopted in the models, as shown by our own results. 
Additionally, we observe no strong evidence of the break in the halo profile 
beyond 100\,kpc reported by \citet{Fukushima2019} using blue horizontal-branch (BHB) stars, nor of the break at $\sim$40\,kpc found by \citet{Keller2008} (based on RRLs), \citet{Akhter2012} (RRLs), and \citet{Thomas2018} (BHB stars). 
Regarding the inner slope values, we observe that our results are higher than those from the literature overall. 
We attribute this to our inability to map the inner halo and contamination in our sample within 20\,kpc (in particular, the misclassification of RRc near the Galactic plane). 
The measured outer slopes from each of our regions are broadly consistent with previous studies based on RRLs, and our value from using all the fields lies within $2\sigma$ of the $n_2$ of most of the works shown in Figure~\ref{fig:comparison}. 
Our values of $n_2$ are also consistent with the SPL density profiles of recent works that focus on tracers at large distances but are unsuccessful at accurately mapping the inner halo (and thus, are unable to find a break).
This is the case of the studies of RRLs and BHB stars by \citet{Medina2018} and \citet{Yu2024}, respectively (which are not shown in Figure~\ref{fig:comparison} for consistency), and most remarkably those of \citet{Lopez-Corredoira2024}'s with K and M giants. 
We note, however, that our density outer slopes are systematically steeper than those from previous works mapping the distant halo ($R_{\rm GC}>80$\,kpc) with BPLs using different halo tracers \citep[e.g.,][]{Thomas2018,Fukushima2019}, and the recent RRL-based SPL outer halo exploration by \citet{Feng2024}. 
This might be caused by contamination in their tracer samples (e.g., by blue stragglers in BHB star catalogues), an incomplete census of RRLs with large distances in our catalogue, or intrinsically different radial distribution for different stellar samples. 
We highlight the similarity of our outer halo results regardless of whether we use RRab-only samples or all RRL types, 
which is likely a consequence of the high number fraction of RRab in our sample.

\subsubsection{Density profiles in context}

The radial density profile of RRLs that we analyse in this work extends to larger radii than any previous study of similar or larger surveyed area. 
Additionally, the extrapolation of our measured slope beyond the 145\,kpc radius used for MCMC modeling provides a good match to the outermost data points in Figure~\ref{fig:dens}, suggesting that the slope we measure is appropriate even in the outermost halo (beyond $\sim$200\,kpc).
While the $\sim$350\,deg$^2$ area covered by our study is much smaller than that used in many other measurements of halo density profiles, our results are consistent with those of most previous studies (see Figure~\ref{fig:comparison}). 
We also note that our sample of RRLs with relatively well-sampled lightcurves is unlikely to suffer significant contamination, and thus constitutes a robust and secure sample of outer halo stars.

\begin{figure*}
\begin{center}
\includegraphics[angle=0, width=0.475\textwidth]{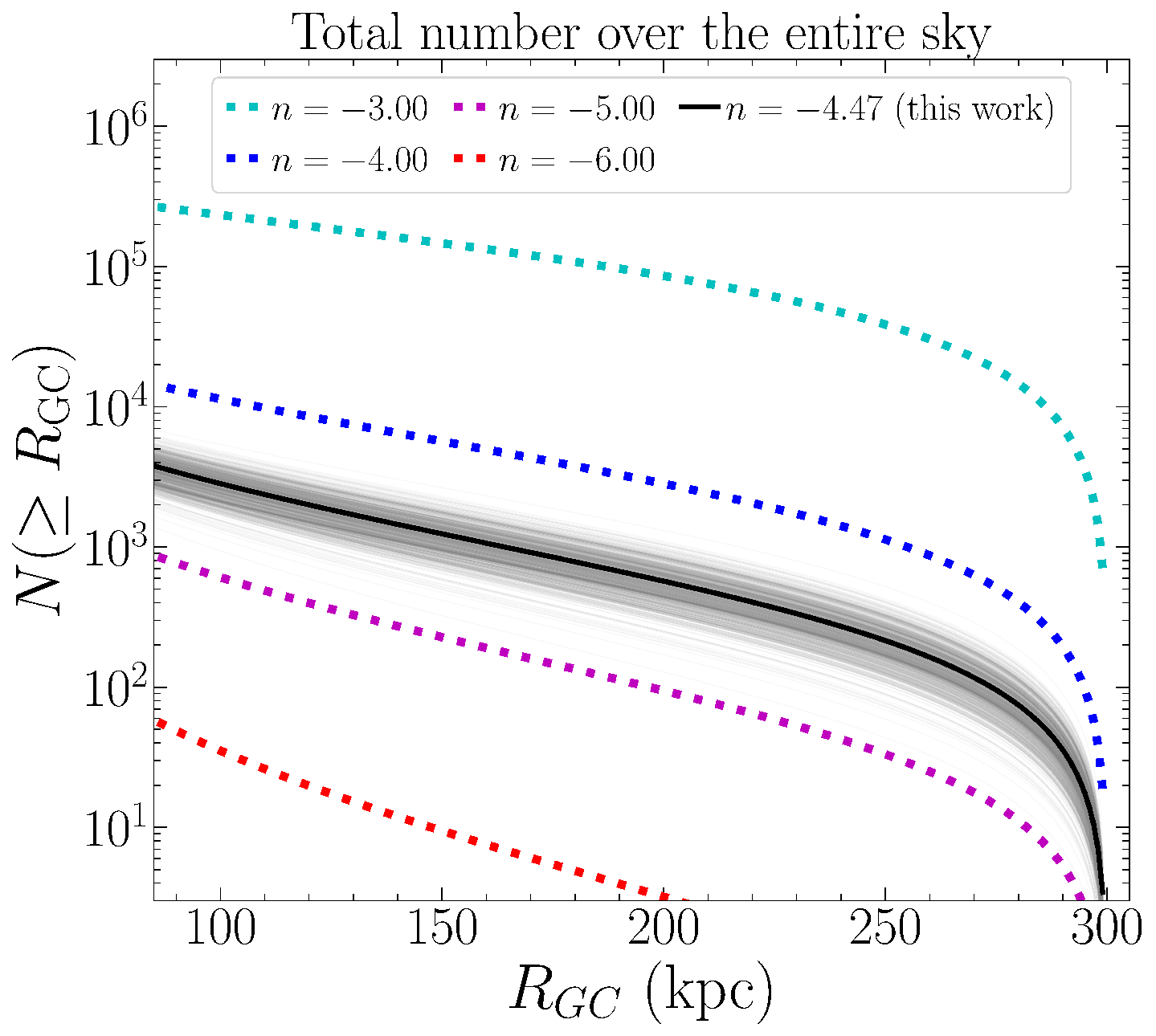}
\includegraphics[angle=0,width=0.475\textwidth]{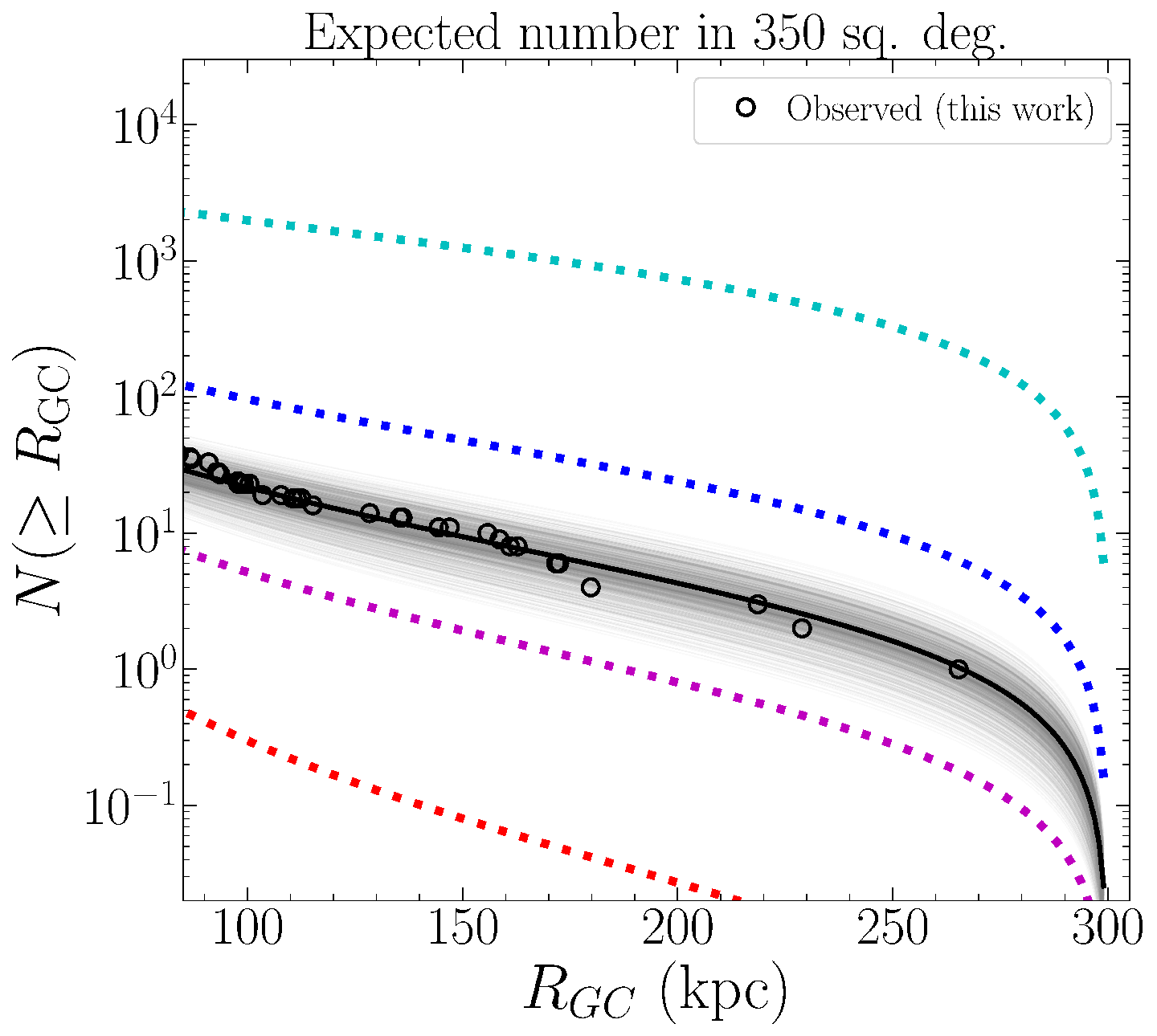}
\caption{
Expected number of RRLs for density profiles with different slopes (for ellipsoidal halo models). 
{\it Left:} Predictions over the entire sky, showing radial density slopes of $n = -3, -4, -5, -6$, and our measured value of $n_2 = -4.47$. 
A total of 
$\sim$3000 RRLs are predicted at a distance larger than 100\,kpc based on our fit, and only $\sim$350 at distances $>230$\,kpc. 
{\it Right:} Predicted number of stars in a survey similar to ours, with area of 350\,deg$^2$. 
Open circles show the actual number of outer halo RRLs analyzed in our study. 
Finding two RRLs beyond 230\,kpc 
rules out the most extreme slopes illustrated here, as a shallow slope (e.g., $n \gtrsim -3.5$) would predict hundreds of stars at these distances and steep slopes (e.g., $n \lesssim -5.5$)  would make it unlikely to find any RRLs at $R_{\rm GC} > 230$\,kpc in a 350\,deg$^2$ survey. 
In both panels, grey lines are used to represent 1000 random drawings from the MCMC chains described in Section~\ref{sec:models}. }
\label{fig:expectations}
\end{center}
\end{figure*}

Figure~\ref{fig:expectations} illustrates the total number of stars beyond a given radius for various power-law outer halo density profiles. 
For our best-fit oblate halo model, $n_2 = -4.47$, a total of $\sim$3000 RRLs are expected over the entire sky beyond 100\,kpc, or roughly 0.073 per square degree 
This is comparable to the model-based expectations from \citealt{Sanderson2017}, which predict $\sim$2000--6000 RRLs from unbound merger remnants beyond 100\,kpc in the MW (see their Figure~5). 
For our survey of 350\,deg$^2$, we would thus expect to find 25 RRLs at $R_{\rm GC} > 100$\,kpc, 
while we find a total of 23.   
Interestingly, the right panel of Figure~\ref{fig:expectations} suggests that the mere identification of two secure RRLs beyond 230\,kpc 
in our 350\,deg$^2$ study rules out outer halo slopes with $n \lesssim -5$ or $n \gtrsim -3.5$. 
Over the entire sky, an outer halo slope of $n = -4.47$ predicts only $\sim$350 RRLs beyond 230\,kpc in the entire MW. 

\begin{table*}
\small
\caption{
Parameters of halo number density profiles from previous works, for simple and broken power law models. 
}
\label{tab:powerLaw}
\begin{center}
\begin{tabular}{|c| c| c| c| c| c| c| c| c| c|}
\hline
\hline
	Model  &  Slope  &  $R_{\rm break}$  &  Inner Slope  &  Outer slope & Range & Tracer Used & Paper\\
	 &    &   (kpc)    &    &    & (kpc) &  & \\
\hline
Simple Power Law    &    &    &   &   &    &  &   \\  
  &  $-3.03\pm 0.08$ &  --  &  --  & -- & $1-80$  & RRLs  &    \citealt{Wetterer1996}\\
  &  $\sim -2.8$ &  --  & -- & -- & 4--60  & RRLs  &   \citealt{Vivas2006}\\
    &  $-3.0$  &  --  &  --  &  --  &  5--40  & MSTO stars & \citealt{Bell2008}  \\
    &  $-2.5\pm 0.2$  &  --  &  --  &  --  &  10--90  & BHB stars & \citealt{DePropris2010}  \\
    &  $-2.7\pm 0.5$  &  --  &  --  &  --  &  $<40$  & BHB stars & \citealt{Deas11}  \\ 
   &  $-2.42\pm0.13$ &  --  & -- & --  & 5--30  & RRab  &  \citealt{Sesar13a}\\
    &  $<-6$   &  --  &  --  &  --  &  50--100  & A-type stars & \citealt{Deas14}  \\ 
    &  $-4.0$  &  --  &  --  &  --  &  50--90  & RRab & \citealt{Cohen2017}  \\    
    &  $-3.4\pm 0.1$  &  --  &  --  &  --  &  10--80  & K giants & \citealt{Xue2015}  \\
    &  $-3.5\pm 0.2$  &  --  &  --  &  --  &  30--90  & Giant stars & \citealt{Slater2016}  \\
    &  $-2.96\pm 0.05$  &  --  &  -- & --  & $<28$  & RRLs & \citealt{Iorio2018}  \\
    &  $-4.17^{+0.18}_{-0.20}$  &  --  &  --  &  --  &  17--145  & RRLs & \citealt{Medina2018}  \\
    &  $-4.09\pm0.10$  &  --  &  --  &  --  &  20--300  & RRLs & \citealt{Feng2024}  \\
    &  $-4.6\pm0.4$  &  --  &  --  &  --  &  25--90  & K giants & \citealt{Lopez-Corredoira2024}  \\
    &  $-4.5\pm0.2$  &  --  &  --  &  --  &  25--90  & M giants & \citealt{Lopez-Corredoira2024}  \\
    &  $-4.28^{+0.12}_{-0.13}$  &  --  &  --  &  --  &  20--70  & BHB stars & \citealt{Yu2024}  \\
\hline
Broken Power Law    &    &    &   &   &    &  &   \\ 
    &  --  &  $45$  &  $-2.48\pm0.09$  &  $-5.0\pm0.3$  &  10--80  & RRLs & \citealt{Keller2008}  \\ 
    &  --  &  $23$  &  $-2.4$  &  $-4.5$  &  5--100  & RRLs & \citealt{Wat09}  \\ 
    &  --  &  $27\pm 1$  &  $-2.3\pm 0.1$  &  $-4.6^{+0.2}_{-0.1}$  &  $10-45$  & BHB and BS stars & \citealt{Deas11}  \\ 
    &  --  &  $28$  &  $-2.6\pm 0.04$  &  $-3.8\pm 0.1$  &  12--40  & near MSTO stars & \citealt{Sesar2011}  \\ 
    &  --  &  $45$  &  $-2.78\pm 0.02$  &  $-5.0\pm 0.2$  &  $10-80$  & RRLs & \citealt{Akhter2012}  \\     
        &  -- &   $26.5\pm8.9$  &  $-2.7\pm 0.6$  &  $-3.6\pm 0.4$  &  $9-49$  & RRLs & \citealt{Faccioli2014}  \\ 
        &  -- &   $24$  &  $-2.8\pm 0.5$  &  $-5.4\pm 0.5$  &  5--60  & RRLs & \citealt{Zinn14}  \\ 
    &  --  &  $20$  &  $-2.5\pm 0.4$  &  $-4.9\pm 0.4$  &  10--60  & F stars & \citealt{PilaDiez2015}  \\ 
    &  --  &  $18\pm 1$  &  $-2.1\pm 0.3$  &  $-3.8\pm 0.1$  &  10--80  & K giants & \citealt{Xue2015}  \\ 
    &  --  &  $29.87^{+2.80}_{-3.55}$  &  $-3.61^{+0.15}_{-0.16}$  &  $-4.75^{+0.30}_{-0.28}$  &  10--70  & BHB stars & \citealt{Das2016}  \\ 
    &  --  &  $41.40^{+2.50}_{-2.40}$  &  $-4.24\pm0.08$  &  $-3.21\pm0.07$  &  $15-220$  & BHB stars & \citealt{Thomas2018}  \\ 
    &  --  &  $160^{+18}_{-19}$  &  $-2.9^{+0.3}_{-0.3}$  &  $-15.0^{+4.5}_{-3.7}$  &  35--360  & BHB stars & \citealt{Fukushima2019}  \\ 
    &  --  &  $32.10^{+1.10}_{-0.90}$  &  $-2.54^{+0.09}_{-0.09}$  &  $-5.42^{+0.13}_{-0.14}$  &  $9-100$  & RRLs & \citealt{Stringer21}  \\

    &  --  &  $31.11^{+7.61}_{-5.88}$  &  $-2.46^{+0.18}_{-0.20}$  &  $-3.99^{+0.75}_{-1.33}$  &  $<80$  & RRLs & \citealt{Chen2023}  \\  
     
    &  --  &  $18.1^{+2.1}_{-1.1}$  &  $-2.05^{+0.13}_{-0.15}$  &  $-4.47^{+0.11}_{-0.18}$  &  7--145  & RRLs & \gm{This work}  \\   
\hline
\end{tabular}
\end{center}
\end{table*}

\section{Discussion and Summary}
\label{sec:discussionSummary}

We have described the search for RRLs in different directions of the remote MW halo using DECam data from the HiTS and the ongoing HOWVAST surveys.
We construct light curves from time series containing from 16 to 38, and from 15 to 32 epochs in the $g$ and $r-$bands, respectively.
Considering all of the studied fields in HiTS and HOWVAST, 
we detect a total of 492 RRLs (397 in the HOWVAST fields and 95 in the HiTS fields) in a combined area of $\sim$270\,deg$^2$, including at least 
90 RRLs not listed in existing surveys. 
When combining our catalog with that of our previous study \citep{Medina2018}, the number of RRLs increases to 663 over a total area of $\sim350$\,deg$^2$. 
The heliocentric distances of our RRLs range between $7$ and $270$\,kpc. 
We identified 9 new RRLs beyond $100$\,kpc, which we add to the still small list of well-characterized tracers of the old component of the MW at large distances.

The distribution of distant RRab stars in the Bailey diagram is not preferentially located in a unique Oo-group, and they are rather uniformly distributed in the period-amplitude space. 
In this subsample, we identified two potential RRL associations containing stars with similar distances (both at around $110$\,kpc) located within a few degrees from each other, and with a mean period of 0.71\,d, consistent with them being linked to the OoII group. 
This might be an indication of them being accreted material from ultra-faint dwarf galaxies. 
Moreover, previous studies have shown that 
neighboring stars at these distances are unlikely to occur by chance \citep[][]{Ses14,BakerWillman2015,Sanderson2017},  which makes them potential tracers of known or undiscovered substructures. 
We found that the position of these groups is inconsistent with those of previously known streams (e.g., the Sagittarius stream), and cannot directly associate these stars with the accretion of UFDs with our data only.
We conclude that the stars in these groups are likely associated, and advocate for additional data and follow-up studies to confirm their association and to determine their parent populations. 

We characterized the (radial) spatial distribution of our RRLs with power-law profiles, by adopting a spherical halo model ($q=1$) and an ellipsoidal model (oblate, with $q=0.7$).  
Furthermore, we analyze the density profiles from the RRLs in our entire sample, and from different directions in the halo.
For this, we followed an MCMC approach and consider RRLs located at distances $<$145\,kpc from the Galactic centre for our modeling.  
We found that the profiles are better described by broken-power-laws, as it has been shown by other works in the literature.
For the preferred model (spherical halo), our best-fitting results suggest a break in the RRLs distribution at $18.1^{+2.1}_{-1.1}$\,kpc, with an inner slope of $-2.05^{+0.13}_{-0.15}$, and a steeper outer slope of $-4.47^{+0.11}_{-0.18}$.

Stellar haloes are important testbeds sensitive to various aspects of galaxy formation models, and comparing observations (e.g., the properties of their density profiles) with simulations is an important requirement to draw  meaningful conclusions. 
In recent years, several authors have measured the stellar distribution of MW-like galaxies using sophisticated cosmological simulation suites \citep[e.g., the IllustrisTNG project;][]{Pillepich2018}.
In these simulations, outer halo slopes are typically found to be in the range $-5.5 < n < -3.5$, where recently formed haloes or those with a large fraction of their total stellar mass originating from mergers have shallower slopes, and steeper slopes correspond to quiescent recent accretion histories \citep{Pillepich2014}. 
Our measured density profiles are remarkably consistent with the results reported by \citet{Merritt2020}, who predicted a median outer slope (beyond 20\,kpc) of $-4.5$ for MW-like galaxies using the TNG100 simulation of the IllustrisTNG project \citep[e.g.,][]{Pillepich2018,Nelson2019}. 

Better number statistics (with large scale halo surveys of high completeness) resulting from the advent of large and deep photometry campaigns will be crucial to reconstruct a more complete local and global picture of the outer halo's history, structure, inclination, and shape. 
These studies will permit us, e.g., to assess the amount of variation in stellar density at large radii \citep[where asymmetries are expected to be more evident; e.g.][]{Pandey2022} and to better characterize the disequilibrium state caused by the dynamical response of the Galactic halo to the infall of massive satellites \citep[observed over thousands of square degrees; see e.g.][]{Conroy2021,Han2022,Rozier2022}. 
In this regard, the upcoming ten-year Rubin Observatory Legacy Survey of Space and Time \citep[LSST;][]{LSST2009} will recover a highly complete sample of the thousands of outer halo RRLs expected out to $\sim$400\,kpc \citep[e.g.,][]{Ivezic08,Oluseyi12,Hernitschek2022}, which will serve as uniquely valuable tools when contrasted with cosmological predictions to disentangle the MW formation in unprecedented detail.


\section*{Acknowledgements}

We thank the anonymous referee for her/his thorough feedback, which helped improve the quality of this manuscript. 
GEM acknowledges support from the University of Toronto Arts \& Science Postdoctoral Fellowship program.   
GEM and EKG gratefully acknowledge the support of the Hector Fellow Academy and the Deutsche Forschungsgemeinschaft (DFG, German Research Foundation) -- Project-ID 138713538 -- SFB 881 (“The Milky Way System”, subproject A03). RRM gratefully acknowledges support by the ANID BASAL project FB210003 and ANID Fondecyt  project 1221695.
JLC acknowledges support from National Science Foundation (U.S.A.) grant AST-1816196. 
CEMV is supported by the international Gemini Observatory, a program of NSF’s NOIRLab, which is managed by the Association of Universities for Research in Astronomy (AURA) under a cooperative agreement with the National Science Foundation, on behalf of the Gemini partnership of Argentina, Brazil, Canada, Chile, the Republic of Korea, and the United States of America. 
CJH has received funding from the European Union’s Horizon 2020 research and innovation programme under grant agreement No 101008324 (ChETEC-INFRA) and the State of Hesse within the Research Cluster ELEMENTS (Project ID 500/10.006). 

This study used data obtained with the Dark Energy Camera (DECam), which was constructed by the Dark Energy Survey (DES) collaboration. Funding for the DES Projects has been provided by the U.S. Department of Energy, the U.S. National Science Foundation, the Ministry of Science and Education of Spain, the Science and Technology Facilities Council of the United Kingdom, the Higher Education Funding Council for England, the National Center for Supercomputing Applications at the University of Illinois at Urbana-Champaign, the Kavli Institute of Cosmological Physics at the University of Chicago, Center for Cosmology and Astro-Particle Physics at the Ohio State University, the Mitchell Institute for Fundamental Physics and Astronomy at Texas A\&M University, Financiadora de Estudos e Projetos, Funda\c{c}\~ao Carlos Chagas Filho de Amparo, Financiadora de Estudos e Projetos, Funda\c{c}\~ao Carlos Chagas Filho de Amparo \`{a} Pesquisa do Estado do Rio de Janeiro, Conselho Nacional de Desenvolvimento Cient\'ifico e Tecnol\'ogico and the Minist\'erio da Ci\^{e}ncia, Tecnologia e Inova\c{c}\~ao, the Deutsche Forschungsgemeinschaft and the Collaborating Institutions in the Dark Energy Survey. The Collaborating Institutions are Argonne National Laboratory, the University of California at Santa Cruz, the University of Cambridge, Centro de Investigaciones En\'ergeticas, Medioambientales y Tecnol\'ogicas–Madrid, the University of Chicago, University College London, the DES-Brazil Consortium, the University of Edinburgh, the Eidgen\"{o}ssische Technische Hochschule (ETH) Z\"{u}rich, Fermi National Accelerator Laboratory, the University of Illinois at Urbana-Champaign, the Institut de Ci\`{e}ncies de l'Espai (IEEC/CSIC), the Institut de F\'isica d'Altes Energies, Lawrence Berkeley National Laboratory, the Ludwig-Maximilians Universit\"{a}t M\"{u}nchen and the associated Excellence Cluster Universe, the University of Michigan, the National Optical Astronomy Observatory, the University of Nottingham, the Ohio State University, the University of Pennsylvania, the University of Portsmouth, SLAC National Accelerator Laboratory, Stanford University, the University of Sussex, and Texas A\&M University.

Based on observations at Cerro Tololo Inter-American Observatory, NSF’s NOIRLab (NOIRLab Prop. ID 2017B-0253 and 2018A-0215; PI: Carlin), which is managed by the Association of Universities for Research in Astronomy (AURA) under a cooperative agreement with the National Science Foundation.
Part of the observations were allocated by the Chilean Telescope Allocation Committee (CNTAC) under the 
Prop. ID 2015A-0608 (PI: F{\"o}rster), 2017B-0907 (PI: Mu\~noz), and 2018A-0907 (PI: Mu\~noz).  

The results of this work were obtained using data from the European Space Agency (ESA) mission {\it Gaia}, processed by the {\it Gaia} Data Processing and Analysis Consortium  (DPAC). 
Funding for the DPAC has been provided by  national institutions, in particular the institutions participating in the {\it Gaia} Multilateral Agreement (MLA).
The {\it Gaia} mission website is \url{https://www.cosmos.esa.int/gaia}. 
The {\it Gaia} archive website is \url{https://archives.esac.esa.int/gaia}.

This research has made use of {\sc pandas} \citep{McKinney10}, {\sc numpy} \citep{vanderWalt11}, the {\sc Astropy} library \citep{Astropy13,Astropy18}, the software {\sc TOPCAT} \citep{Taylor05}, and the VizieR catalogue access tool, CDS, Strasbourg, France. 
The original description of the VizieR service was published in A\&AS 143, 23.
The figures in this paper were produced with {\sc Matplotlib } \citep{Hunter07}.

\section*{Data Availability}

The data underlying this article are available as online supplementary material. 
Additional data from this work will be shared upon reasonable request to the corresponding author.

\newpage 

\appendix

\section{Complementary material}

\begin{figure*}
\begin{center}
\includegraphics[angle=0,scale=.30]{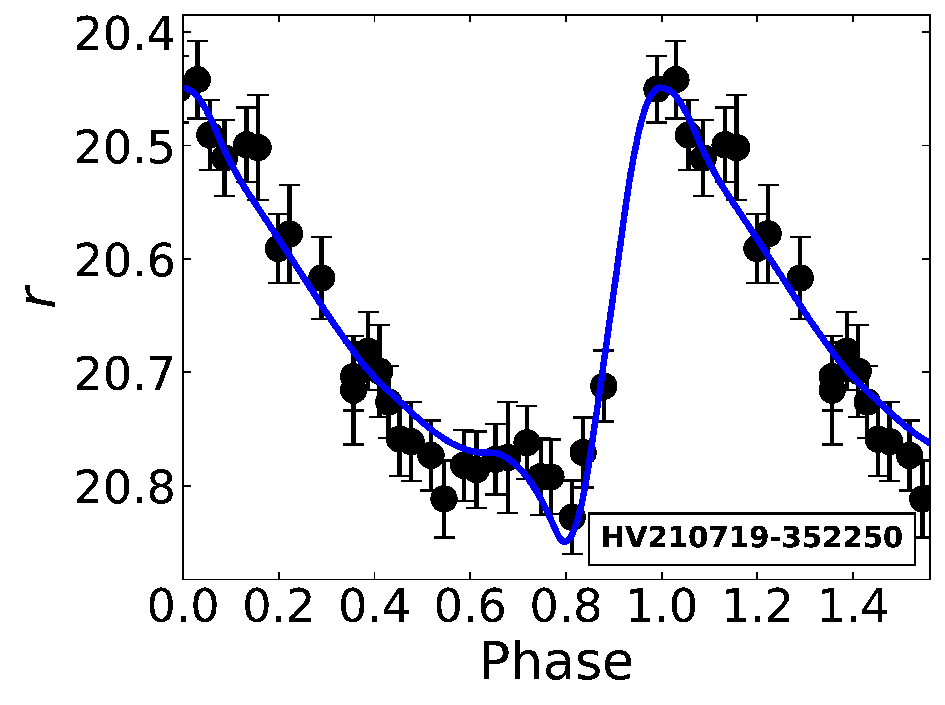}
\includegraphics[angle=0,scale=.30]{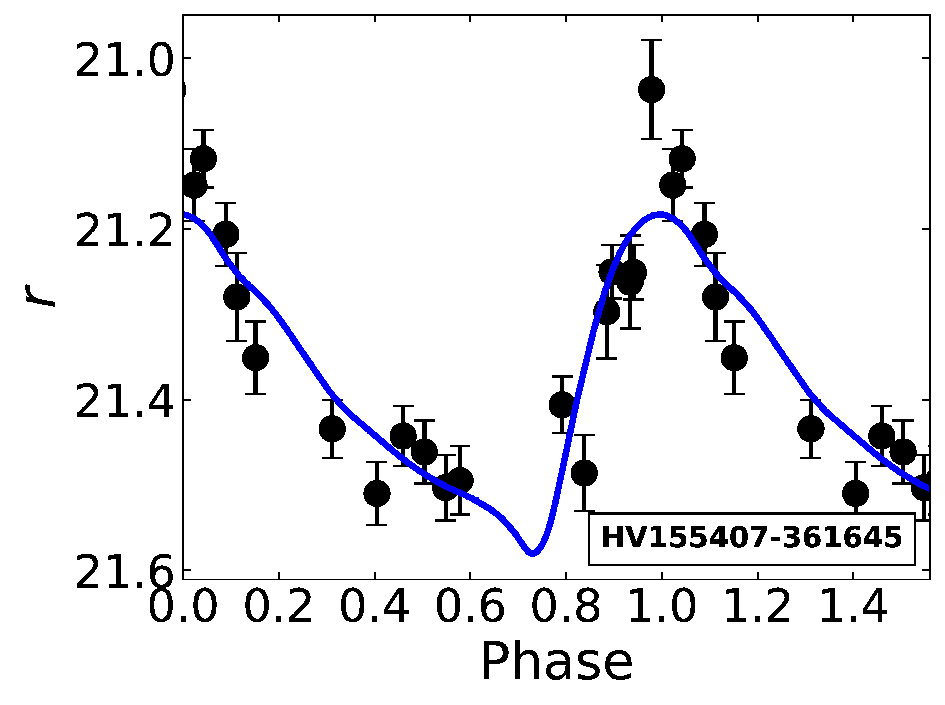}
\includegraphics[angle=0,scale=.30]{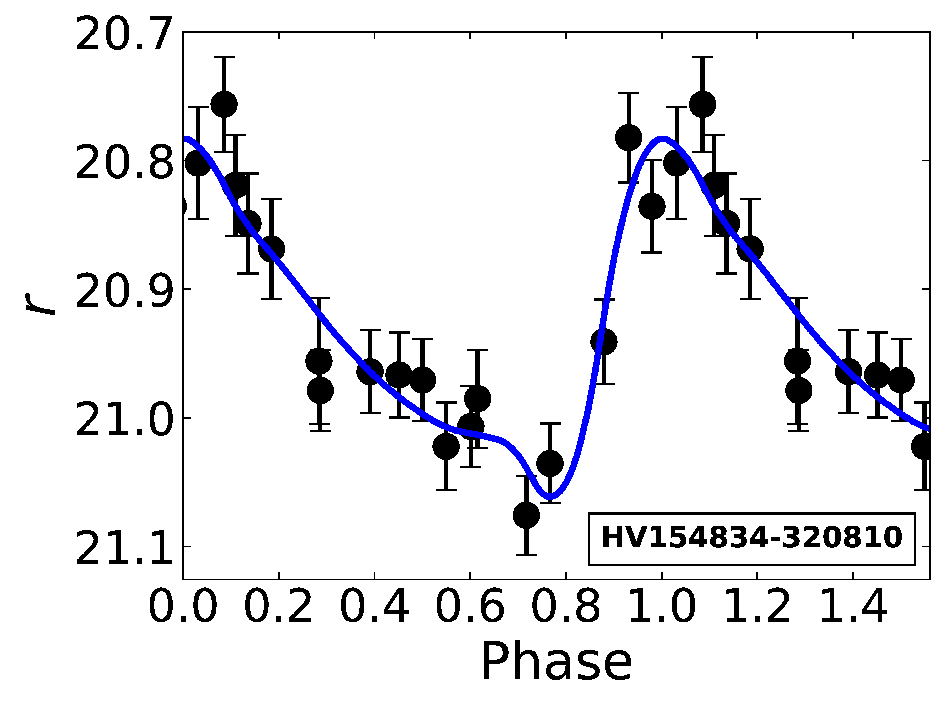}\\
\includegraphics[angle=0,scale=.30]{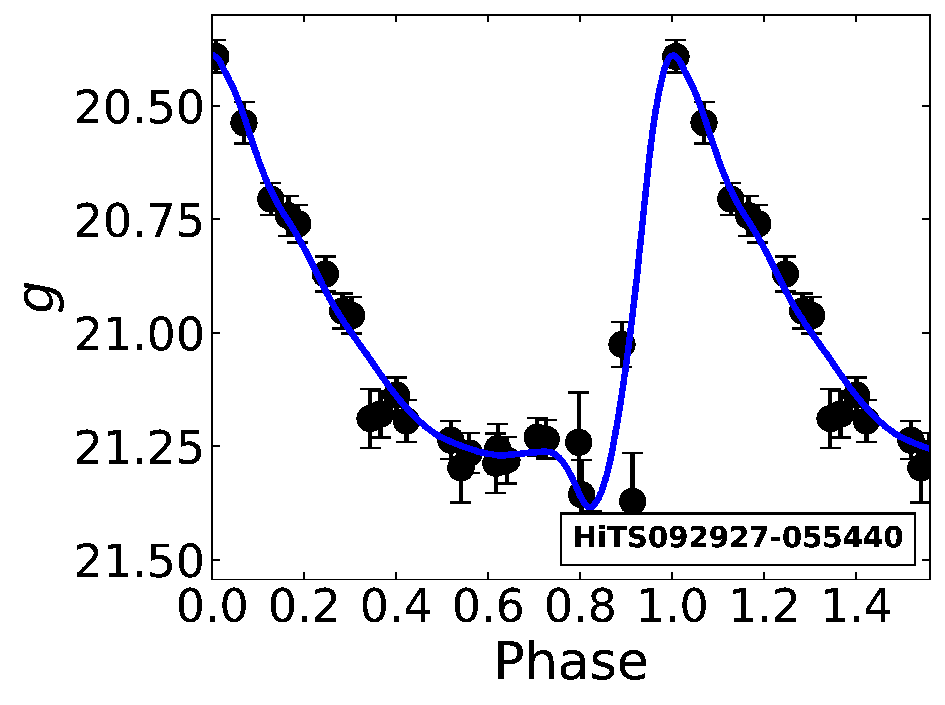}
\includegraphics[angle=0,scale=.30]{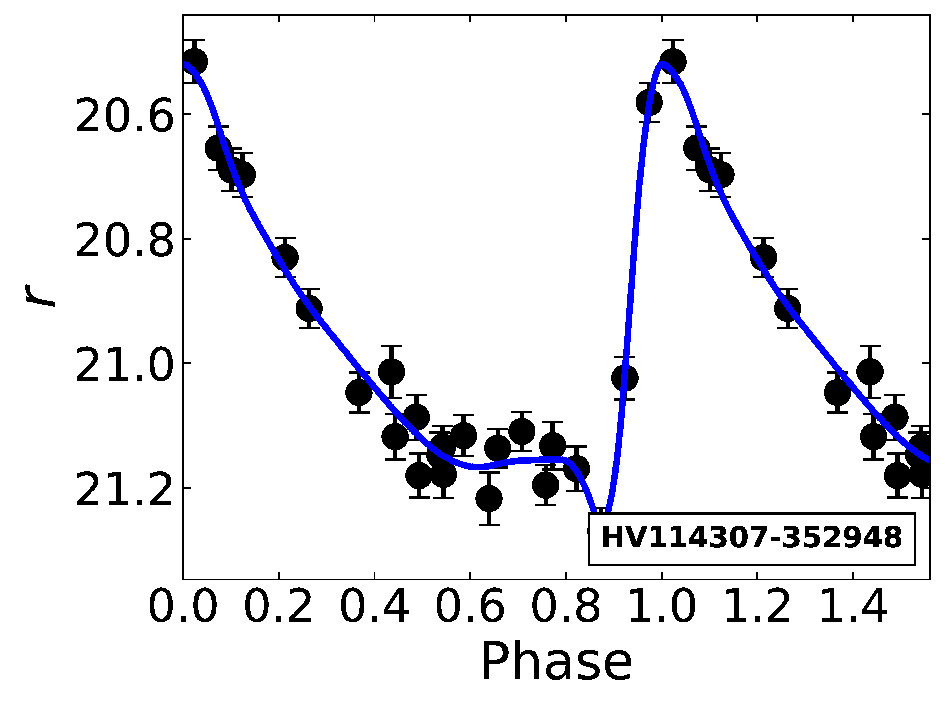}
\includegraphics[angle=0,scale=.30]{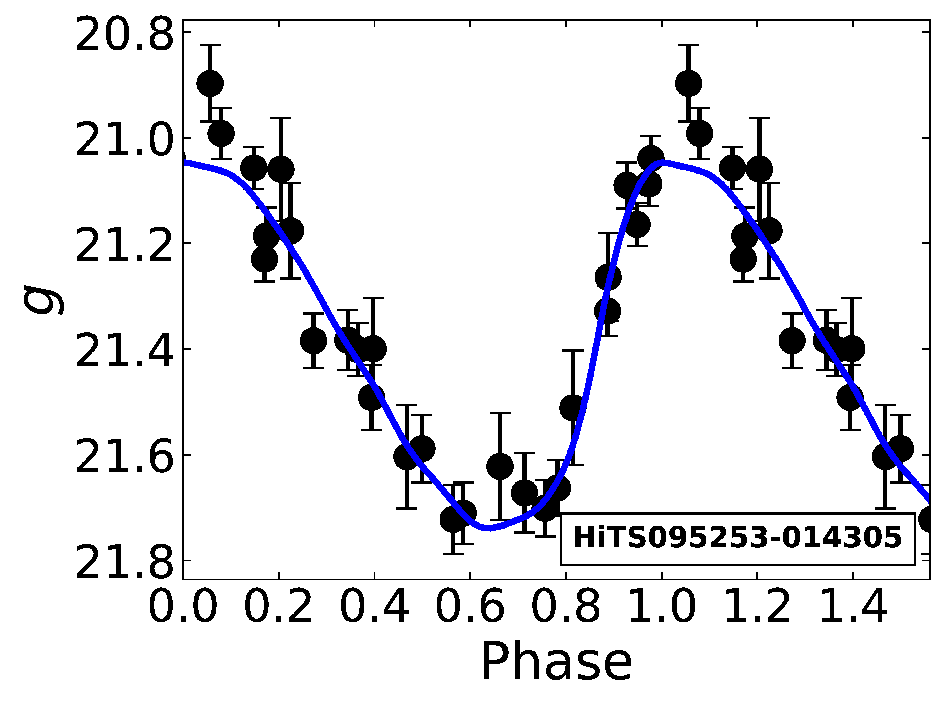}\\
\includegraphics[angle=0,scale=.30]{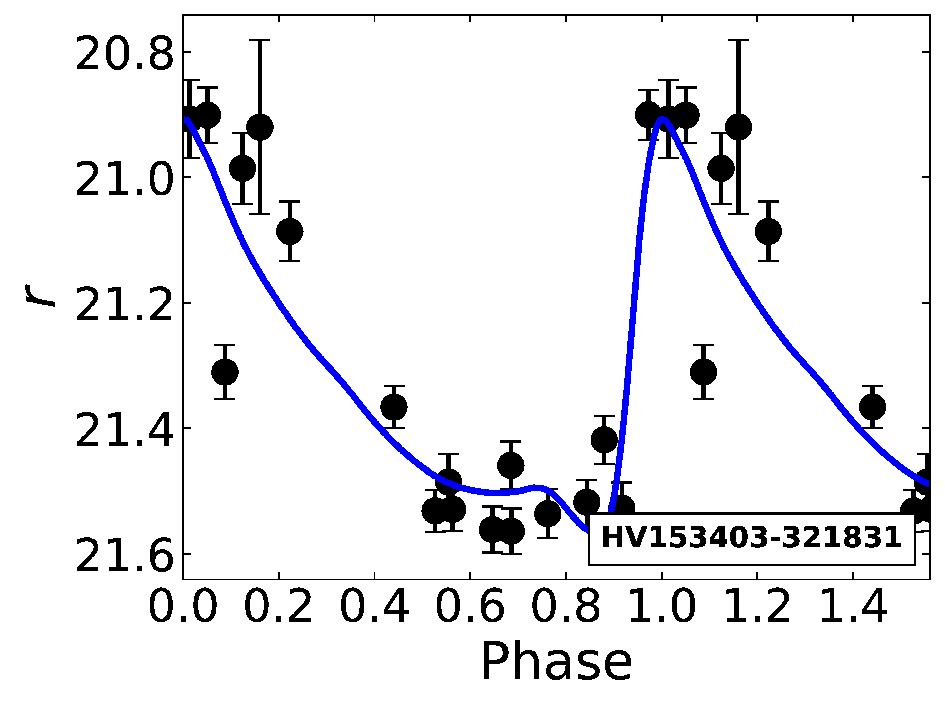}
\includegraphics[angle=0,scale=.30]{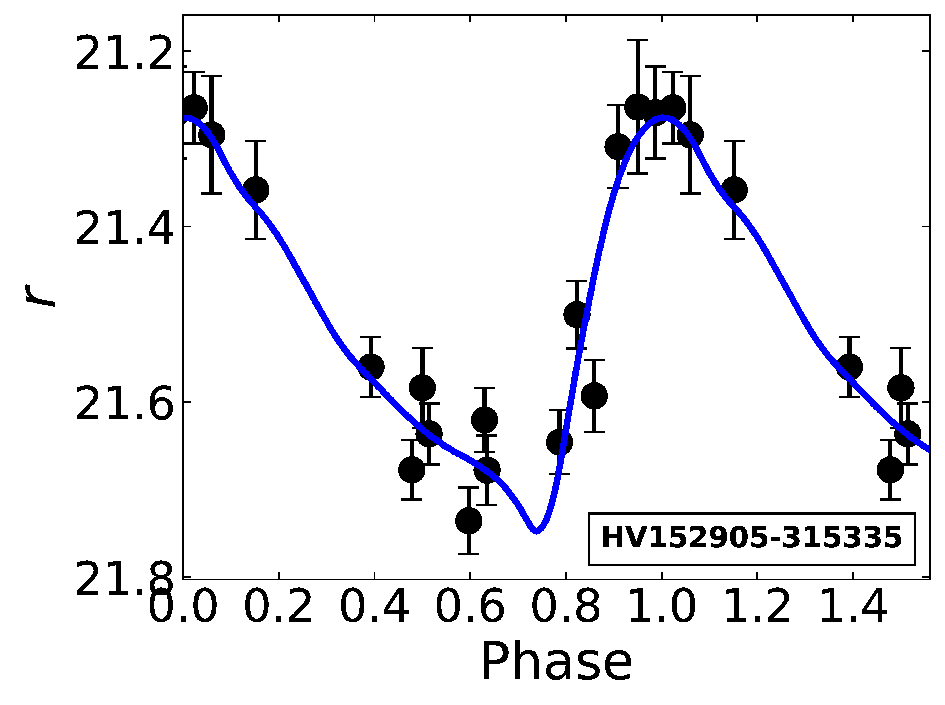}
\includegraphics[angle=0,scale=.30]{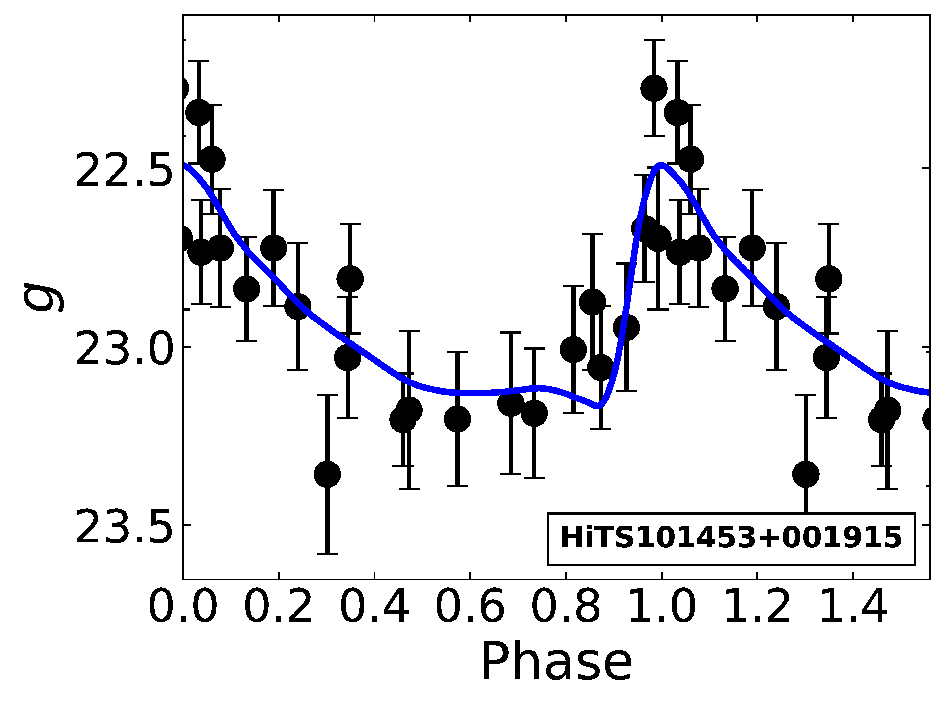}
\caption{
Folded light curves in the $g$ and $r$ band of our sample of RRLs with $d_{\rm H}>100$\,kpc, and out to $\sim$270\,kpc.
For each star, we overplot the best-fitting model from the Python module \textsc{gatspy} \citep{VanderPlas2015} with a blue solid line. 
These models were obtained from the SDSS Stripe 82 RRLs light curve templates \citep{Sesar2010}.
The main properties of these RRLs are summarized in Table~\ref{tab:distant}.
}
\label{fig:lcs}
\end{center}
\end{figure*}

\begin{figure*}
\begin{center}
\includegraphics[angle=0,scale=.45]{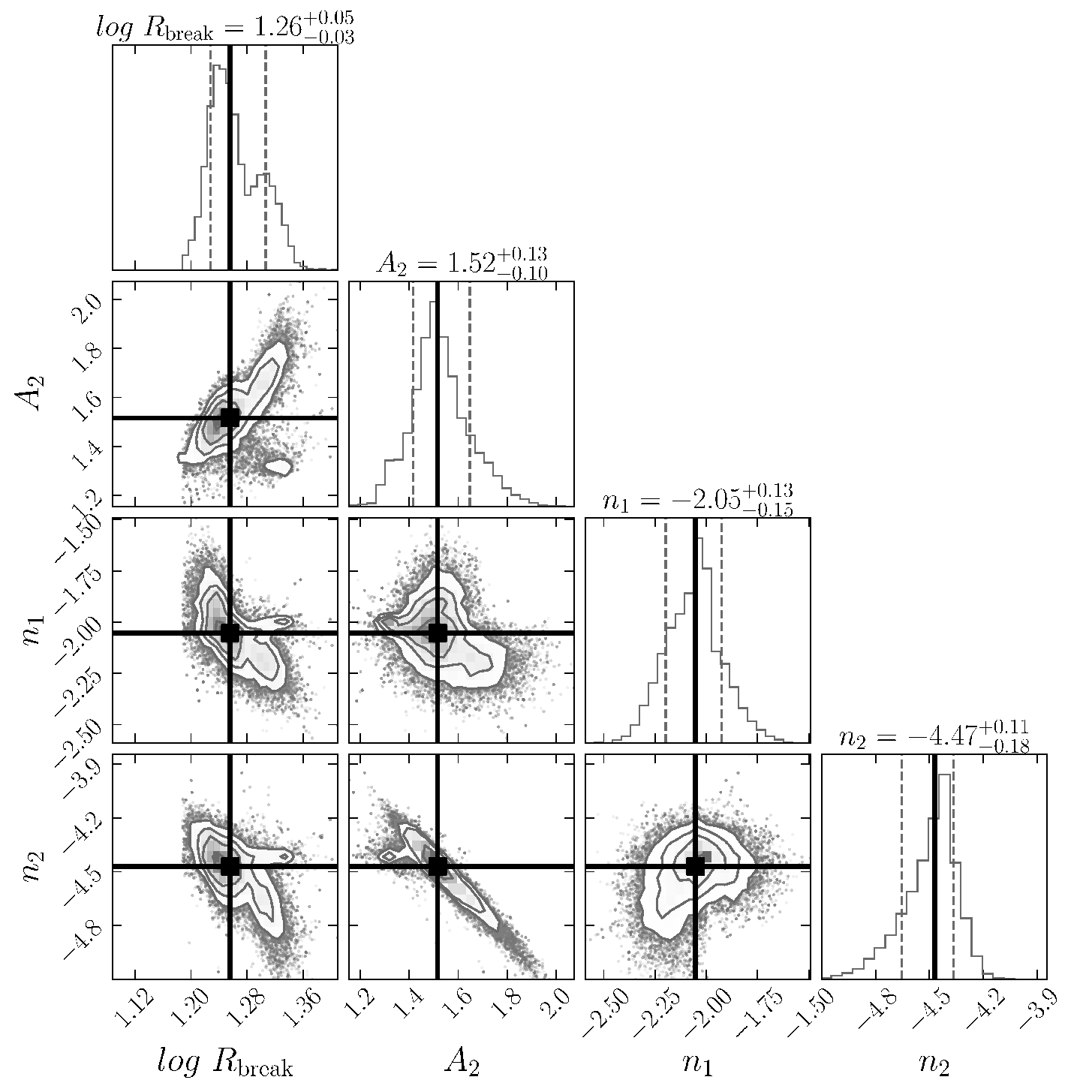}
\caption{
Corner plot of the posterior probability distributions for the broken-power-law profiles described in Section~\ref{sec:density}.
The adopted value of each parameter is the median of the corresponding marginalized distribution, and their uncertainties represent the 16th and 84th percentiles. 
These parameters are computed for the RRLs in all the fields studied in this work. }
\label{fig:corner1}
\end{center}
\end{figure*}

In this section, we provide additional material to complement the content of this manuscript. 
More specifically, Table~\ref{tab:fulltable} lists the full list of RRLs detected in this work, 
summarizing their main properties. 
In Figure~\ref{fig:lcs}, we display the light curves of the most distant RRLs in our sample (beyond 100\,kpc).  
Finally, we depict the marginalized parameter distribution (as corner plots) resulting from our MCMC analysis when using our entire RRL sample (Figure~\ref{fig:corner1}).

\begin{table*}
\small
\caption{
Same as Table~\ref{tab:distant} but for the entire sample of RRLs analysed in this work. 
The full table is provided as supplementary material and will be made available at the Strasbourg astronomical Data Center
(CDS). 
}
\label{tab:fulltable}
\begin{center}

\begin{tabular}{|c|c|c|c|c|c|c|c|c|c|c|c|}
\hline
ID &           RA &          DEC &  $<g>$ &           $<r>$ & $N_g$ &  $N_r$ & Period$^*$ & Amplitude$^*$ & Type &            $d_{\rm H}$ & Flag\\
  & (deg) & (deg) &   &  & & & (days ) &  & & (kpc)  &    \\
\hline
 HiTS090839-003849 &  $137.16200$ &   $-0.64690$ &  $17.05$ &  $17.21$ &  $25$ &    -- &  $0.3689$ &    $0.48$ &    c &    $18.7\pm0.7$ &  $0$ \\
 HiTS091013-050321 &  $137.55230$ &   $-5.05570$ &  $15.66$ &  $15.78$ &  $26$ &    -- &  $0.5412$ &    $1.24$ &   ab &     $9.2\pm0.3$ &  $0$ \\
 HiTS091047+015033 &  $137.69500$ &    $1.84240$ &  $18.30$ &  $18.05$ &  $27$ &    -- &  $0.6021$ &    $0.94$ &   ab &    $32.2\pm1.2$ &  $0$ \\
 HiTS091050-055917 &  $137.70940$ &   $-5.98800$ &  $19.45$ &  $19.39$ &  $28$ &    -- &  $0.6468$ &    $0.47$ &   ab &    $55.6\pm2.1$ &  $0$ \\
 HiTS091110-062237 &  $137.78990$ &   $-6.37700$ &  $18.39$ &  $18.29$ &  $28$ &    -- &  $0.5713$ &    $0.85$ &   ab &    $32.5\pm1.2$ &  $0$ \\
 HiTS091139-003904 &  $137.91340$ &   $-0.65120$ &  $16.48$ &  $16.60$ &  $25$ &    -- &  $0.4742$ &    $1.33$ &   ab &    $12.8\pm0.5$ &  $0$ \\
 HiTS091156+022530 &  $137.98310$ &    $2.42490$ &  $16.45$ &  $16.29$ &  $27$ &    -- &  $0.6021$ &    $1.32$ &   ab &    $13.7\pm0.5$ &  $0$ \\
 HiTS091510-052952 &  $138.79140$ &   $-5.49790$ &  $20.63$ &  $20.49$ &  $28$ &    -- &  $0.6199$ &    $0.97$ &   ab &    $93.6\pm3.5$ &  $0$ \\
 HiTS091512+021915 &  $138.80090$ &    $2.32080$ &  $17.70$ &  $17.97$ &  $28$ &    -- &  $0.2855$ &    $0.65$ &    c &    $23.2\pm1.0$ &  $0$ \\
 HiTS091528-041929 &  $138.86600$ &   $-4.32460$ &  $16.46$ &  $16.63$ &  $24$ &    -- &  $0.3872$ &    $0.62$ &    c &    $14.5\pm0.5$ &  $0$ \\
 
\hline

\end{tabular} 
\end{center}
$^*$The period and amplitude of pulsation are computed from the photometric band with more observations. 
\end{table*}

\bsp	
\label{lastpage}
\end{document}